# Liquidity Adjustment in Multivariate Volatility Modeling: Evidence from Portfolios of Cryptocurrencies and US Stocks

Qi Deng[1,2,*]

## Abstract

We develop a liquidity-sensitive multivariate volatility framework to improve the estimation of time-varying covariance structures under market frictions. We introduce two novel portfolio-level liquidity measures, liquidity jump and liquidity diffusion, which capture magnitude and volatility of liquidity fluctuation, respectively, and construct liquidity-adjusted return and volatility that reflect real-time liquidity variability. These liquidity-adjusted inputs are integrated into a VECM-DCC/ADCC-Bayesian model, allowing for conditional and posterior covariance estimation under liquidity stress. Applying this framework to portfolios of cryptocurrencies and US stocks, we find that traditional models misrepresent volatility and co-movement, while liquidity-adjusted models yield more stable and interpretable risk structures, particularly for portfolios of cryptocurrencies. The findings support the use of liquidity-adjusted multivariate models as statistically grounded tools for assessing the propagation of portfolio risk under market frictions, with implications for asset pricing, market microstructure design, and portfolio management.

Funding Source: The work was supported by Hubei University of Automotive Technology [grant number BK202209] and Hubei Provincial Bureau of Science and Technology [grant number 2023EHA018].

1. College of Artificial Intelligence, Hubei University of Automotive Technology, Shiyan, China
2. Cofintelligence Financial Technology Ltd., Hong Kong and Shanghai, China
*. Corresponding author: dq@huat.edu.cn; qi.deng@cofintelligence.com

**Liquidity Adjustment in Multivariate Volatility Modeling: Evidence from Portfolios of Cryptocurrencies and US Stocks**

**1. Introduction**

Modeling multivariate return dynamics under conditions of liquidity variability presents a core challenge in empirical asset pricing and portfolio optimization. Financial markets frequently exhibit time-varying liquidity conditions, including episodic shocks, shifting bid-ask spreads, and fluctuations in market depth. These frictions disrupt the assumptions of multivariate econometric models such as vector autoregression (VAR) and dynamic conditional correlation (DCC), which typically assume normality and stationarity in asset return dynamics. As a result, these models often fail to accurately capture co-movement structures and underestimate risk in environments with high liquidity variability. These limitations are especially pronounced in emerging asset classes, most notably cryptocurrencies, where liquidity risk is both persistent and extreme. Cryptocurrencies exhibit highly fragmented market structures, abrupt changes in trading volume, and frequent disruptions in price discovery, all of which contribute to unstable and discontinuous return dynamics. Liquidity distortions also undermine the autoregressive structure of returns, making conventional multivariate time series models unreliable in forecasting both return and volatility dynamics, and impairing portfolio risk estimation. While the literature has recognized the role of liquidity in asset pricing, there remains no comprehensive framework to adjust return and volatility inputs at the portfolio level using real-time liquidity information to restore the modeling and forecasting power of theses multivariate econometric models.

This paper develops a liquidity-sensitive modeling framework that adjusts both return and volatility with real-time liquidity conditions, and provides empirical evidence that the liquidity adjustment significantly enhances the stability and predictability of multivariate volatility models. Specifically, we construct portfolio-level liquidity-adjusted return and volatility that serve as inputs to the multivariate econometrics and portfolio optimization models. In addition, we introduce two daily portfolio-level liquidity metrics: portfolio liquidity jump (capturing the magnitude of discrete liquidity shocks) and portfolio liquidity diffusion (capturing the variation in liquidity variation across time). These liquidity measures and adjustments enable a more robust characterization of the true underlying return-generating process, allowing econometric models to better align with actual liquidity conditions.

The liquidity-adjusted return and volatility and the portfolio liquidity measures form the foundation for all subsequent modeling process by capturing how liquidity conditions reshape both return and risk structures. We use the liquidity-adjusted return and volatility as inputs to a flexible and extensible econometric structure: a vector error correction model (VECM) combined with dynamic conditional correlation (DCC) and asymmetric DCC (ADCC) specifications. The integration of liquidity-adjusted return and volatility into this VECM-DCC/ADCC structure improves the estimation of both long-run equilibrium relationships and short-run volatility dynamics. Furthermore, we extend the forecasting power of the VECM-DCC/ADCC model by combining a Bayesian posterior covariance updating process. We demonstrate that liquidity-adjusted return and volatility lead to more stable correlation dynamics and improved predictive performance across varying liquidity regimes. The unified VECM-DCC/ADCC-Bayesian framework can be applied to both regular and liquidity-adjusted series. We demonstrate that the predictive power and risk estimation accuracy of the framework is significantly enhanced when it is applied to liquidity-adjusted return and volatility, especially for assets that exhibit extreme liquidity variations, such as cryptocurrencies.

We operationalize the VECM-DCC/ADCC-Bayesian framework in the context of classical mean-variance (MV) portfolio optimization, constructing a series of Liquidity-adjusted Mean-Variance (LAMV) portfolios with different levels of covariance estimation. The resulting Liquidity-Adjusted Mean-Variance (LAMV) model incorporates both liquidity-sensitive return expectations and Bayesian-updated covariance matrices. We conduct empirical tests to compare the performance of LAMV portfolios against that of traditional mean-variance (TMV) portfolios across two distinct asset classes: cryptocurrencies and US stocks. These asset classes differ sharply in their liquidity profiles, providing a robust testbed for assessing the effectiveness of liquidity-adjusted modeling and evaluating the generalizability of the LAMV model. Our empirical results reveal several key findings. First, liquidity adjustment significantly enhances the stability and predictive performance of multivariate volatility models for cryptocurrencies. Second, the LAMV portfolios outperform TMV portfolios across both asset classes, with especially pronounced gains in cryptocurrencies, of which the liquidity variation is more extreme. Third, even sophisticated forecasting models such as DCC/ADCC and Bayesian shrinkage offer limited improvements unless return and volatility are adjusted for liquidity. For US stocks, where liquidity is more stable, the LAMV portfolios still yield meaningful performance improvements, underscoring the broader relevance of the framework.

We contribute to the literature on liquidity, market microstructure and asset co-movement in three ways. First, we introduce a multivariate modeling framework that explicitly accounts for trading frictions by embedding liquidity jump and diffusion measures into a VECM-DCC/ADCC-Bayesian structure. These inputs allow us to estimate covariance matrices that evolve with liquidity fluctuations and capture frictions that are commonly observed but rarely incorporated into multivariate risk modeling. Second, we provide evidence that liquidity-adjusted volatility and correlation estimates differ substantially from those obtained using traditional models. These differences are most pronounced in assets with extreme liquidity variability, such as cryptocurrencies, but are also observed in more liquid asset classes during stress periods. Third, we show that these liquidity-induced changes in co-movement structure carry implications for pricing and portfolio allocation. By incorporating liquidity directly into statistical modeling rather than treating it as a portfolio constraint, we provide a new modeling paradigm that extends conventional multivariate volatility models to better reflect liquidity-driven risk, contributing to ongoing debates about liquidity risk, volatility clustering, and cross-market dynamics.

By centering the modeling process on liquidity-adjusted fundamentals as opposed to relying solely on structural econometric sophistication, this study provides a practical and scalable modeling framework for portfolios of assets with high liquidity variations, and advances a more accurate, flexible, and adaptive approach to dynamically allocate these assets. The proposed liquidity adjustment methodology and the Bayesian-enhanced multivariate autoregressive framework have direct implications for asset pricing, volatility forecasting, and portfolio allocation for assets characterized by high liquidity risk, offering useful and practical tools to investors, risk managers, and policymakers seeking to navigate liquidity-volatile asset classes.

The rest of the paper proceeds as follows. Section 2 reviews existing literature on portfolio liquidity. Section 3 introduces portfolio liquidity jump and diffusion, and portfolio-level liquidity-adjusted return and volatility. Section 4 provides descriptive statistics of the dataset and discussions on the distributions of portfolio-level liquidity measures. Section 5 presents the VECM-DCC/ADCC-Bayesian model and forecasts of posterior covariance matrix. Section 6 optimizes LAMV portfolios enhanced with forecasts from the liquidity-adjusted VECM-DCC/ADCC-Bayesian models. Section 7 concludes the paper.

## 2. Literature Review on Portfolio Liquidity

We refer to Deng and Zhou (2024, 2024) for a comprehensive review of asset-level liquidity measures, liquidity costs, components of liquidity, and models for assets with extreme liquidity risk. In this section, we focus on recent developments in portfolio liquidity modeling in the context of portfolio optimization.

In the literature, liquidity has often been incorporated as a constraint in portfolio optimization. Among the vast body of portfolio optimization research (see, e.g., Kolm, Tütüncü and Fabozzi, 2014, for a review), only a select subset explicitly includes liquidity constraints. Lo, Petrov and Wierzbicki (2006) define portfolio liquidity as the weighted average liquidity of individual assets, coining the term Weighted Average Liquidity (WAL). Vieira and Filomena (2019) consider the total monetary value of a portfolio (Financial Value Liquidation or FVL) and model liquidity with parameters reflecting the practices of portfolio managers. More recently, Vieira et al. (2023) employ both WAL and FVL to examine how liquidity constraints affect index tracking, finding that portfolios with liquidity constraints are more liquid than unconstrained portfolios.

On the methodological side, a considerable stream of literature seeks closed-form solutions to dynamic portfolio optimization with liquidity costs. These include models with transient liquidity impact (e.g., Çetin and Rogers, 2007; Ly Vath, Mnif and Pham, 2007; Ma, Song and Zhang, 2013) and those with permanent market impact (e.g., Gârleanu and Pedersen, 2013; Lim and Wimonkittiwat, 2014; Gaigi et al., 2016; Mei, DeMiguel and Nogales, 2016). To address cases where closed-form solutions are intractable, semi closed-form and numerical approaches like the Least-Squares Monte Carlo (LSMC) algorithm have been applied to dynamic portfolio optimization (Brandt et al., 2005; Cong and Oosterlee, 2016, 2017; Zhang et al., 2019) to broaden the applicability of dynamic optimization in the presence of trading frictions and liquidity costs.

Another line of studies examines liquidity's impact on portfolio Value-at-Risk (VaR), again often treating liquidity as a constraint, which is more directly relevant to our methodology. Al Janabi (2011) argues that regular VaR models assess the downward risk in mark-to-market portfolio value over a given time horizon but do not account for the actual trading risk of liquidation and introduces a multivariate Liquidity-Adjusted VaR (LVaR) subject to constraints on expected return, trading volume and liquidation horizon. Al Janabi (2013) extends this by incorporating a GARCH-M(1,1) component (volatility feedback) into a multivariate LVaR, effectively blending time-varying volatility and expected return forecasts into the liquidity-constrained VaR construct. Hung et al. (2020) apply multivariate GARCH-t and GJR-GARCH-t models to incorporate liquidity properties embedded in individual asset returns, evaluating how these models improve LVaR forecast accuracy. Weiß and Supper (2013) model the joint distribution of bid-ask spreads and log returns of a stock portfolio by using Autoregressive Conditional Double Poisson and GARCH processes and vine copulas. Al Janabi, Ferrer and Shahzad (2019) develop a LVaR optimization technique based on vine copulas for multi-asset portfolios. Al Janabi et al. (2017) propose a nonlinear DCC *t*-copula model to replace linear correlations in LVaR computation. Al Janabi (2021) provides a thorough review on LVaR-based multivariate portfolio optimization algorithms.

On the other hand, liquidity has seldom been used a direct parameter in portfolio optimization. At the theoretical level, the existing literature does not model portfolio-level (and for that matter, asset-level) return and volatility with explicit adjustment of liquidity. We aim to bridge this gap by providing models that explicitly adjust return and volatility with liquidity, which serves as our first motivation. At the methodological level, although a few studies provide certain autoregressive models that address portfolio-level liquidity risk (e.g., Al Janabi, 2013; Weiß and Supper, 2013; Al Janabi et al., 2017; Hunt et al., 2020), again these models do not incorporate explicit liquidity-adjustment on the variables they actually model (i.e., portfolio-level conditional return and covariance); they treat liquidity as an external constraint or a parallel process. In this paper, we aim to fill this gap by developing a new set of liquidity-adjusted multivariate autoregressive models specifically designed for portfolios of assets with extreme liquidity variability, with improved predictability. This serves as our second motivation.

## 3. Liquidity-adjusted Return and Volatility, and Liquidity Measures

### 3.1 Liquidity-adjusted Return and Volatility, and Liquidity Measures on Asset Level

In this subsection, we briefly review the asset-level liquidity jump and liquidity diffusion proposed by Deng and Zhou (2024), who model the minute-level liquidity-adjusted volatility ${\sigma^2}_T^\ell$ and return $r_t^\ell$ for time-period $T$ (a 24-hour/1440-minute trading day for cryptocurrencies, or 6.5-hour/390-minute trading day for US stocks) as follows ($\tau$ as minute-level time index):

$${\sigma^2}_T^\ell = \frac{1}{T}\sum_{\tau=1}^{T} \eta_T \frac{|r_\tau|/\overline{|r_\tau|}}{A_\tau/\overline{A_\tau}} (r_\tau - \bar{r}_\tau)^2 = \frac{1}{T}\sum_{\tau=1}^{T}\left(r_\tau^\ell - \overline{r_\tau^\ell}\right)^2$$

$$r_\tau^\ell = \sqrt{\eta_T \frac{|r_\tau|/\overline{|r_\tau|}}{A_\tau/\overline{A_\tau}}} r_\tau$$

$$subject\ to: \sum_{\tau=1}^{T} \eta_T \frac{|r_\tau|/\overline{|r_\tau|}}{A_\tau/\overline{A_\tau}} = T \Rightarrow \eta_T = \frac{T}{\sum_{\tau=1}^{T} \frac{|r_\tau|/\overline{|r_\tau|}}{A_\tau/\overline{A_\tau}}}$$

*where*:
1. *$r_\tau$ is the observed return at minute $\tau$, $|r_\tau|$ is its absolute value, $\overline{|r_\tau|}$ is its arithmetic average in that day,*
2. *$A_\tau$ is the dollar amount traded at minute $\tau$, $\overline{A_\tau}$ is its arithmetic average in day T,*
3. *$\eta_T$ is the daily normalization factor on day T and is a constant for day T,*
4. *T=1440 or 390 as there are 1440 minutes (24 hours) or 390 minutes (6.5 hours) in a crypto or US stock trading day*

The daily (day-level) regular and liquidity-adjusted returns for time-period $T$ is obtained by aggregating the intraday (minute-level) returns are given as ($t$ as daily time index):

$$r_t = (1 + r_\tau)^T - 1$$
$$r_t^\ell = (1 + r_\tau^\ell)^T - 1$$

The realized and unobservable daily (intraday on minute-level) variance for time-period $T$ is:

$$\sigma^{2\ell}_t = T\sigma^{2\ell}_T$$

Deng and Zhou (2025a) then define a "daily liquidity Beta" pair: a "liquidity jump" that measures the magnitude of daily liquidity fluctuation, and a "liquidity diffusion" that measures the daily (intraday) liquidity volatility. The "liquidity jump" $\beta_{r_t}^\ell$, and the "liquidity diffusion" $\beta_{\sigma_t}^\ell$ are defined as follows[1]:

$$\beta_{r_t}^\ell = |r_t/r_t^\ell| \subset \begin{cases} > 1; high\ daily\ liquidity\ fluctuation \\ = 1; equilibrium\ daily\ liquidity\ fluctuation \\ < 1; low\ daily\ liquidity\ fluctuation \end{cases} \qquad (1)$$

$$\beta_{\sigma_t}^\ell = \sigma_t/\sigma_t^\ell \subset \begin{cases} > 1; high\ daily\ liquidity\ volatility \\ = 1; equilibrium\ daily\ liquidity\ volatility \\ < 1; low\ daily\ liquidity\ volatility \end{cases} \qquad (2)$$

The liquidity jump $\beta_{r_t}^\ell$ captures the sudden and discontinuous change of daily liquidity level of an asset. A low-liquidity asset (e.g., a cryptocurrency) would have higher $\beta_{r_t}^\ell$ values, indicating its high liquidity fluctuation, while a high-liquidity asset (e.g., a US stock) would have lower $\beta_{r_t}^\ell$ values. The liquidity diffusion $\beta_{\sigma_t}^\ell$ reflects the intraday volatility of liquidity conditions, that an asset with stable liquidity (trading volume), such as a US stock, would have lower $\beta_{\sigma_t}^\ell$ values, while an asset with high liquidity, such as a cryptocurrency, would have higher $\beta_{\sigma_t}^\ell$ values.

**3.2 Liquidity-adjusted Return and Volatility, and Liquidity Measures on Portfolio Level**

---

[1] It is possible that $r_t/r_t^\ell$ is negative, therefore $\beta_{r_t}^\ell$, a positive value, is defined as $|r_t/r_t^\ell|$.

Using the same notations as Deng and Zhou (2024), at the portfolio level (with $N$ assets, $N > 1$), the daily regular and liquidity-adjusted return vectors ($Q_t$ and $Q_t^\ell$, respectively), and how they are connected, are given as follows:

$$Q_t = \begin{bmatrix} r_{t_1} \\ \vdots \\ r_{t_i} \\ \vdots \\ r_{t_N} \end{bmatrix}; \quad Q_t^\ell = \begin{bmatrix} r_{t\,1}^\ell \\ \vdots \\ r_{t\,i}^\ell \\ \vdots \\ r_{t\,N}^\ell \end{bmatrix}; \quad i \in [1, N] \tag{3}$$

$$Q_t = \begin{bmatrix} r_{t_1} \\ \vdots \\ r_{t_i} \\ \vdots \\ r_{t_N} \end{bmatrix} = \begin{bmatrix} \beta_{r_{t_1}}^\ell r_{t\,1}^\ell \\ \vdots \\ \beta_{r_{t_i}}^\ell r_{t\,i}^\ell \\ \vdots \\ \beta_{r_{t_N}}^\ell r_{t\,N}^\ell \end{bmatrix} = \begin{bmatrix} \beta_{r_{t_1}}^\ell & \cdots & 0 & \cdots & 0 \\ \vdots & \cdots & \vdots & \cdots & \vdots \\ 0 & \cdots & \beta_{r_{t_i}}^\ell & \cdots & 0 \\ \vdots & \cdots & \vdots & \cdots & \vdots \\ 0 & \cdots & 0 & \cdots & \beta_{r_{t_N}}^\ell \end{bmatrix} \begin{bmatrix} r_{t\,1}^\ell \\ \vdots \\ r_{t\,i}^\ell \\ \vdots \\ r_{t\,N}^\ell \end{bmatrix} = B_{r_t}^{\mathcal{P}\ell} Q_t^\ell \Rightarrow Q_t^\ell = B_{r_t}^{\mathcal{P}\ell^{-1}} Q_t \tag{4}$$

where:

$$B_{r_t}^{\mathcal{P}\ell} = \begin{bmatrix} \beta_{r_{t_1}}^\ell & \cdots & 0 & \cdots & 0 \\ \vdots & \cdots & \vdots & \cdots & \vdots \\ 0 & \cdots & \beta_{r_{t_i}}^\ell & \cdots & 0 \\ \vdots & \cdots & \vdots & \cdots & \vdots \\ 0 & \cdots & 0 & \cdots & \beta_{r_{t_N}}^\ell \end{bmatrix} \tag{5}$$

In Equation 5, we introduce a "portfolio liquidity jump matrix" $B_{r_t}^{\mathcal{P}\ell}$, which is a diagonal matrix with elements being asset-level $\beta_{r_{t_i}}^\ell$'s. The portfolio-level regular and liquidity-adjusted daily covariance matrices, and their connection, are given as follows:[2]

$$\Sigma_{r_t}^{TT} = T \begin{bmatrix} \sigma_{t\,11}^2 & \cdots & \sigma_{t\,1i} & \cdots & \sigma_{t\,1N} \\ \vdots & \cdots & \vdots & \cdots & \vdots \\ \sigma_{t\,i1} & \cdots & \sigma_{t\,ii}^2 & \cdots & \sigma_{t\,iN} \\ \vdots & \cdots & \vdots & \cdots & \vdots \\ \sigma_{t\,N1} & \cdots & \sigma_{t\,Ni} & \cdots & \sigma_{t\,NN}^2 \end{bmatrix}; \quad \Sigma_{r_t^\ell}^{TT} = T \begin{bmatrix} \sigma_{t\,11}^{2\,\ell} & \cdots & \sigma_{t\,1i}^\ell & \cdots & \sigma_{t\,1N}^\ell \\ \vdots & \cdots & \vdots & \cdots & \vdots \\ \sigma_{t\,i1}^\ell & \cdots & \sigma_{t\,ii}^{2\,\ell} & \cdots & \sigma_{t\,iN}^\ell \\ \vdots & \cdots & \vdots & \cdots & \vdots \\ \sigma_{t\,N1}^\ell & \cdots & \sigma_{t\,Ni}^\ell & \cdots & \sigma_{t\,NN}^{2\,\ell} \end{bmatrix} \tag{6}$$

$$\Sigma_{r_t}^{TT} = B_{\sigma_t}^{\mathcal{P}\ell} \Sigma_{r_t^\ell}^{TT} B_{\sigma_t}^{\mathcal{P}\ell\,H} \tag{7}$$

where: $\sigma_{t_{ij}}$ is the intraday (minute-level) covariance between assets $i$ and $j$, $i, j \in [1, N], i \neq j$

---

[2] From this point forward we use the phrase "daily covariance matrices" to refer to the daily (minute-level intraday) covariance matrices $\Sigma_{r_t}^{TT}$ and $\Sigma_{r_t^\ell}^{TT}$ to avoid being verbose.

In Equation 7 we introduce a "portfolio liquidity diffusion matrix" $B_{\sigma_t}^{\mathcal{P}\ell}$, which is a symmetric matrix and can be solved by a Conditional Singular Value Decomposition (Conditional SVD) method proposed by Deng (2024) (see Appendix 1 for details). Collectively $B_{r_t}^{\mathcal{P}\ell}$ and $B_{\sigma_t}^{\mathcal{P}\ell}$ form the "portfolio liquidity matrix pair." In addition, the determinant of $B_{r_t}^{\mathcal{P}\ell}$, $|B_{r_t}^{\mathcal{P}\ell}|$, is regarded as the "portfolio liquidity jump," a scalar that measures the portfolio liquidity fluctuation (size); the determinant of $B_{\sigma_t}^{\mathcal{P}\ell}$, $|B_{\sigma_t}^{\mathcal{P}\ell}|$, is the "portfolio liquidity diffusion" scalar that measures the portfolio liquidity volatility. Their significance becomes apparent in Sections 4 and 5.

## 4. Dataset and Descriptive Statistics

We collect trading data for two portfolios with very different liquidity profiles: one portfolio of cryptocurrencies (characterized by lower overall liquidity and higher liquidity variability) and one portfolio of US stocks (characterized by high liquidity and low liquidity variability). For each portfolio, we construct daily returns and covariances (regular and liquidity-adjusted) and compute the portfolio-level liquidity jump and diffusion measures introduced above. We then examine the distribution of these measures to understand the liquidity dynamics of each portfolio.

### 4.1 Dataset and Descriptive Statistics – cryptocurrency portfolio

First, we select the eight largest non-stable-coin cryptocurrencies by market capitalization with at least six years of historical data (January 17, 2019 to March 7, 2025 with 2,242 trading days) for a five-year back-tests (January 17, 2020 to March 7, 2025, 1,877 trading days) with a 1-year (365 days) rolling window that trade on Binance, the largest cryptocurrency exchange.[3] The selected crypto assets are ADA, BNB, BTC, ETC, ETH, LINK, LTC and XRP. We aggregate the tick-level trading data of these assets to construct and calculate daily measures for both asset and portfolio levels. We aggregate tick-level trading data (transaction records) for each of these cryptocurrencies into one-minute intervals to compute intraday measures. From these, we calculate daily asset-level liquidity-adjusted returns and volatilities, and then aggregate to the portfolio level as described in Section 3, as well as the portfolio liquidity jump $\left|B_{r_t}^{\mathcal{P}\ell}\right|$ and portfolio liquidity diffusion $\left|B_{\sigma_t}^{\mathcal{P}\ell}\right|$. We report the descriptive statistics of $\left|B_{r_t}^{\mathcal{P}\ell}\right|$ and $\left|B_{\sigma_t}^{\mathcal{P}\ell}\right|$ in Panel A of Table 1, and provide their histograms in Column A of Figure 1.

---

[3] https://coinmarketcap.com/rankings/exchanges/, accessed on March 4, 2025.

In this subsection, we specifically discuss the descriptive statistics of portfolio liquidity jump ($\left|B_{r_t}^{\mathcal{P}\ell}\right|$) and portfolio liquidity diffusion ($\left|B_{\sigma_t}^{\mathcal{P}\ell}\right|$). For the cryptocurrency portfolio, from Panel A of Table 1 "liquidity jump $\left|B_{r_t}^{\mathcal{P}\ell}\right|$" column, the mean of $\left|B_{r_t}^{\mathcal{P}\ell}\right|$ is at 3.77 and the median is at 0.70, indicating a highly right-skewed distribution with a long right tail. The number of days with extreme liquidity jump ($\left|B_{r_t}^{\mathcal{P}\ell}\right| = 10$) is 682 (30.42% of 2,242), the number of days with high liquidity jump ($\left|B_{r_t}^{\mathcal{P}\ell}\right| \geq 1$) is 1,051 (46.88% of 2,242), and the number days with low liquidity jump ($\left|B_{r_t}^{\mathcal{P}\ell}\right| \leq 0.10$) is 748 (33.36% of 2,242). As such, there are extremely liquidity jumps on both ends. In Panel A of Table 1 "liquidity diffusion $\left|B_{\sigma_t}^{\mathcal{P}\ell}\right|$" column, the mean and median are 1.38 and 0.33 respectively, again indicating a highly right-skewed distribution with a long right tail. The numbers of days with extreme and high portfolio liquidity diffusion ($\left|B_{\sigma_t}^{\mathcal{P}\ell}\right| = 10, \left|B_{\sigma_t}^{\mathcal{P}\ell}\right| \geq 1$) are 204 (23.42%) and 811 (36.17%), respectively, and the number of days with low portfolio liquidity diffusion ($\left|B_{\sigma_t}^{\mathcal{P}\ell}\right| \leq 0.10$) is 3 (0.13%). These statistics indicate that both liquidity fluctuation and volatility at the portfolio level are very high. The histograms of $\left|B_{\sigma_t}^{\mathcal{P}\ell}\right|$

and $\left|B_{r_t}^{\mathcal{P}\ell}\right|$ visualize their distributions. We observe that there are both extreme liquidity jump ($\left|B_{r_t}^{\mathcal{P}\ell}\right| \gg 1$) and extreme liquidity diffusion ($\left|B_{\sigma_t}^{\mathcal{P}\ell}\right| \gg 1$).

**4.2 Dataset and Descriptive Statistics – stock portfolio**

Second, we collect minute-level trading data of all 1,503 constituent stocks of the SP500 (large cap), SP400 (mid cap) and SP600 (small cap) indices from the Polygon.io API. For each index, we pick five largest stocks in terms of market cap, thus we select 15 stocks from the three indices.[4] All 15 stocks have at least ten years of complete historical data (July 28, 2014 to March 10, 2025 with 2,671 trading days), for back-tests (July 14, 2015 to March 10, 2025, 2,429 trading days) with a 1-year (242 days) rolling window. The selected stocks, in alphabetical order, are AAPL, AMZN, ATI, CMA, CRS, EME, GOOG, IBKR, LII, MLI, MSFT, NVDA, TPL, VFC and WSO. We aggregate the minute-level trading data of these assets to construct and calculate asset-level daily data, from which we construct the return vectors and portfolio covariance matrices, both regular ($Q_t$ and $\Sigma_{r_t}^{TT}$) and liquidity-adjusted ($Q_t^{\ell}$ and $\Sigma_{r_t^{\ell}}^{TT}$), as well as the portfolio liquidity jump and diffusion matrices $B_{r_t}^{\mathcal{P}\ell}$ and $B_{r_t}^{\mathcal{P}\ell}$. We report the descriptive statistics of $\left|B_{r_t}^{\mathcal{P}\ell}\right|$ and $\left|B_{\sigma_t}^{\mathcal{P}\ell}\right|$ in Panel B of Table 1, and provide the histograms of $\left|B_{r_t}^{\mathcal{P}\ell}\right|$ and $\left|B_{\sigma_t}^{\mathcal{P}\ell}\right|$ in Column B of Figure 1.

From Panel B of Table 1 "liquidity jump $\left|B_{r_t}^{\mathcal{P}\ell}\right|$" column, we observe that the distribution of $\left|B_{r_t}^{\mathcal{P}\ell}\right|$ is markedly different from that of the cryptocurrency portfolio; it is heavily concentrated at the low end. The mean is 0.11, the median is 0.00. The number of days with extreme liquidity jump ($\left|B_{r_t}^{\mathcal{P}\ell}\right| = 10$) is very low at 16 (0.60% of 2,671), the number of days with high liquidity jump ($|B_{r_t}^{\mathcal{P}\ell}| \geq 1$) is low at 48 (1.80% of 2,671), and the number days with low liquidity jump ($|B_{r_t}^{\mathcal{P}\ell}| \leq 0.10$) is 2,531 (94.76% of 2,671). As $|B_{r_t}^{\mathcal{P}\ell}|$ represents portfolio liquidity jump, the results indicate the liquidity fluctuation at the portfolio level is very low. In practical terms, this means the liquidity of the stock portfolio is extremely stable day-over-day, and liquidity rarely deteriorates sharply from one day to the next.

---

[4] We obtain the market cap data of all 1,503 stocks from the Yahoo Finance API on October 12, 2024 at market close.

In Panel B of Table 1 "liquidity diffusion $|B_{\sigma_t}^{\mathcal{P}\ell}|$" column, both the mean and median of $|B_{\sigma_t}^{\mathcal{P}\ell}|$ are effectively zero at $\sim o(10^{-2})$, and the numbers of days with extreme and high portfolio liquidity diffusion ($|B_{\sigma_t}^{\mathcal{P}\ell}| = 10, |B_{\sigma_t}^{\mathcal{P}\ell}| \geq 1$) are zero, and the number of days with low portfolio liquidity diffusion ($|B_{\sigma_t}^{\mathcal{P}\ell}| \leq 0.10$) is 2,671 (100%). As $|B_{\sigma_t}^{\mathcal{P}\ell}|$ represents portfolio liquidity diffusion, the portfolio-level liquidity volatility is extremely low, a reflection of the deep and continuous liquidity for stocks. The histograms of $|B_{r_t}^{\mathcal{P}\ell}|$ and $|B_{\sigma_t}^{\mathcal{P}\ell}|$ in Column B of Figure 1 confirm the above findings with visualization.

### 4.3 Comparisons between Stock and Cryptocurrency Portfolios

In this subsection, we specifically discuss the differences of the descriptive statistics of portfolio liquidity jump ($|B_{r_t}^{\mathcal{P}\ell}|$) and portfolio liquidity diffusion ($|B_{\sigma_t}^{\mathcal{P}\ell}|$) between the stock and cryptocurrency portfolios, as they are directly related to the performance of the regular and liquidity-adjusted VECM-DCC-Bayesian models for both asset classes.

We find there are sharp contrasts between the portfolios in the values of portfolio liquidity jump ($|B_{r_t}^{\mathcal{P}\ell}|$) and portfolio liquidity diffusion ($|B_{\sigma_t}^{\mathcal{P}\ell}|$). We observe that the cryptocurrency portfolio has much higher liquidity variability (i.e., orders-of-magnitude higher liquidity fluctuation and volatility) than the stock portfolio, with highly asymmetric distributions for both measures. These

comparisons vindicate that the highly-liquid US stocks have very low liquidity variability while the illiquid cryptocurrencies have very high liquidity variability at the portfolio level. The results are consistent with and extend the findings of Deng and Zhou (2024) at the asset level to the portfolio context: stable and mature assets maintain steady liquidity, while emerging and fragmented assets experience frequent liquidity disruptions. This contrast in liquidity profiles between the two asset classes provides a crucial backdrop for our modeling in the next sections. We expect that the benefits of liquidity adjustment in modeling will be far more pronounced for the cryptocurrency portfolio than for the stock portfolio, given these underlying differences.

## 5. Multivariate Autoregressive Framework and Posterior Covariance Matrix Estimation

Having constructed liquidity-adjusted return and covariance inputs, in this section we present a multivariate time-series framework for modeling and forecasting portfolio covariance matrices. The framework is divided into three stages and is applied in two versions (without and with liquidity adjustment). The three components of the model are: (1) a VECM/VAR($p$) that models the portfolio conditional return, (2) a DCC/ADCC(1,1) that models the portfolio conditional covariance matrix, and (3) a Bayesian update that estimates the posterior covariance matrix. Below we detail this framework for the case of regular (unadjusted) returns and volatility, and then discuss the parallel implementation for liquidity-adjusted data.

### 5.1 VECM-DCC/ADCC-Bayesian Framework for Regular Return and Volatility

Following Deng (2018), we first apply the VECM/VAR model to the daily return series of the stock portfolio containing the 15 selected stocks over the sample period of 2,671 days (with 2,429 days being out-of-sample predictions and a rolling window of 242 days), and that of the cryptocurrency portfolio consisting the eight selected cryptocurrencies over the sample period of 2,242 days (with 1,877 days being out-of-sample predictions and a rolling window of 365 days). We conduct a set of Johansen tests on the full sample and confirm that the $r_t$ of all assets are cointegrated, and so as the $r_t^\ell$.[5] Therefore, both return vectors ($Q_t$ and $Q_t^\ell$) can be modelled by an autoregressive VECM construct. We then establish a specific time series VECM($p$) specification (see Appendix 2 Subsection A2.1 for details) given in Equation 8. For each rolling window, we identify the VECM order $p$ based on the AIC value (with an upper bound of $p \leq 5$), fit the data, and use the residue vector as the fitting error ($E_t$) for the next-stage DCC/ADCC analysis. The VECM($p$) specification of Equation 8 produces a one-period-ahead ($t+1$) forecasted return vector $\hat{Q}_{t+1}$, of which the residue error vector $\hat{E}_{t+1}$ is given in Equation 9 (see Appendix 2 Subsection A2.1 for more details).

Second, we apply both a DCC(1,1) and an ADCC(1,1) specifications to estimate the time-varying conditional covariance matrix in the error vector $\hat{E}_{t+1}$. Our approach is inspired by Ling and McAleer (2003). For each rolling window, we fit both DCC(1,1) (Engle, 2002) and ADCC(1,1) (Cappiello, Engle and Sheppard, 2006) of Equation 10 on $\hat{E}_{t+1}$. The reason we fit ADCC is to allow for different responses to positive vs. negative shocks. We choose either DCC(1,1) or ADCC(1,1) (whichever with higher log-likelihood) to produce the time-varying conditional covariance matrix, $\hat{\Omega}_{t+1}$ (steps of deriving $\hat{\Omega}_{t+1}$ are given in Appendix 2, Subsection A2.1). In essence, this step captures how correlation evolve over time in the context of multivariate autoregression in volatility. For example, if assets have recently experienced shocks, the conditional covariance will adjust upward (volatility and correlations increase), and then decay towards a long-run average if no further shocks occur. The ADCC specification allows correlations to respond differently when there are negative shocks (which often increase correlations disproportionally, a common feature in financial markets).

---

[5] For the purpose of being concise, we do not report the results of Johansen tests in this paper. These results are available upon request.

Finally, with the forecasted conditional covariance matrix $\hat{\Omega}_{t+1}$ we estimate the posterior (forecasted) covariance matrix for day $t+1$, $\hat{\Sigma}^{TT}_{r_{t+1}}$ in Equation 11, which is essentially a Bayesian shrinkage formula with prior information in $\Sigma^{TT}_{r_t}$ (steps of deriving the $\hat{\Sigma}^{TT}_{r_{t+1}}$ are given in Appendix 2 Subsection A2.1).[6] The idea is to improve the robustness of the covariance estimate by shrinking extreme values). Intuitively, if the conditional covariance forecast is very volatile or based on limited data, the Bayesian update pulls it closer to a central estimate (reducing extreme risk forecasts); if the forecast is on solid ground, the adjustment is minor. The end result is the best estimate of the covariance matrix for day $t+1$, incorporating both historical dynamics and a Bayesian smoothing.

The 3-stage process is consolidated /;by Equations 8-11 for the regular VECM-DCC/ADCC-Bayesian framework (see Appendix 2 Subsection A2.1 for details):

$$Q_t = \sum_{i=1}^{p} \Phi_i Q_{t-i} + E_t \qquad (8)$$

$$\hat{E}_{t+1} = \hat{Q}_{t+1} - Q_{t+1} \qquad (9)$$

$$\hat{E}_{t+1}|\Psi_t \sim N(0, \hat{\Omega}_{t+1}) \qquad (10)$$

$$\hat{\Sigma}^{TT}_{r_{t+1}} = \Sigma^{TT}_{r_t} + \left[\left(\tau \Sigma^{TT}_{r_t}\right)^{-1} + \hat{\Omega}^{-1}_{t+1}\right]^{-1} \qquad (11)$$

Where:
1) $Q_t$ is the portfolio return vector, $E_t$ is the residual vector, and $\Phi_i$ is the coefficient matrix for VAR lag i,
2) $\hat{Q}_{t+1}$ is the forecasted portfolio return vector (out-of-sample) at time t+1,
3) $Q_{t+1}$ is the actual observed return vector (out-of-sample) at time t+1,
4) $\hat{E}_{t+1}$ is the conditional residual vector (out-of-sample) at time t+1, from the VECM/VAR stage,
5) $\hat{\Omega}_{t+1}$ is the conditional covariance matrix of $\hat{E}_{t+1}$ in the rolling window

---

[6] We use $\Sigma^{TT}_{r_t}$ and $\Sigma^{TT}_{r^\ell_t}$ to refer to the regular and liquidity-adjusted daily covariance matrices, respectively, to avoid the confusion with $\Sigma_{r_t}$ and $\Sigma_{r^\ell_t}$, which are the regular and liquidity-adjusted covariance matrices for a given period of time (a 242-day rolling window for stocks or a 365-day rolling window for cryptocurrencies), respectively.

6) $\tau$ is a scaling factor.

## 5.2 VECM-DCC/ADCC-Bayesian Framework for Liquidity-Adjusted Return and Volatility

We repeat the procedure in Subsection 5.1 to derive the liquidity-adjusted VECM-DCC/ADCC-Bayesian framework in Equations 12-19 (see Appendix 2 Subsection A2.2 for details):

$$Q_t^\ell = \sum_{i=1}^p \Phi_i Q_{t-i}^\ell + E_t^\ell \tag{12}$$

$$\hat{E}_{t+1}^\ell = \hat{Q}_{t+1}^\ell - Q_{t+1}^\ell \tag{13}$$

$$\hat{E}_{t+1}^\ell | \Psi_t \sim N(0, \hat{\Omega}_{t+1}^\ell) \tag{14}$$

$$\hat{\Sigma}_{r_{t+1}^\ell}^{TT} = \Sigma_{r_t^\ell}^{TT} + \left[\left(\tau \Sigma_{r_t^\ell}^{TT}\right)^{-1} + \hat{\Omega}_{t+1}^{\ell\,-1}\right]^{-1} \tag{15}$$

where:

$$Q_t = B_{r_t}^{\mathcal{P}\ell} Q_t^\ell \Rightarrow Q_t^\ell = B_{r_t}^{\mathcal{P}\ell\,-1} Q_t \tag{16}$$

$$E_t^\ell = B_{r_t}^{\mathcal{P}\ell\,-1} E_t \tag{17}$$

$$\Sigma_{r_t^\ell}^{TT} = B_{\sigma_t}^{\mathcal{P}\ell\,-1} \Sigma_{r_t}^{TT} B_{\sigma_t}^{\mathcal{P}\ell\,H\,-1} \tag{18}$$

$$\hat{\Omega}_{t+1}^\ell = B_{r_t}^{\mathcal{P}\ell\,-\frac{1}{2}} \hat{\Omega}_{t+1} B_{r_t}^{\mathcal{P}\ell\,-\frac{1}{2}} \tag{19}$$

## 5.3 Effect of Liquidity Adjustment on Conditional Covariance Estimate

An important outcome of the framework is understanding how liquidity adjustment affects the estimated conditional covariance. Equation 19 links the liquidity-adjusted and regular conditional covariance matrices, $\hat{\Omega}_{t+1}^\ell$ and $\hat{\Omega}_{t+1}$, in which the portfolio liquidity jump $\left|B_{r_t}^{\mathcal{P}\ell}\right|$ is a scaling factor. When it is high ($\left|B_{r_t}^{\mathcal{P}\ell}\right| > 1$), $\hat{\Omega}_{t+1}^\ell$ is scaled down relative to $\hat{\Omega}_{t+1}$, and when the liquidity jump is low ($\left|B_{r_t}^{\mathcal{P}\ell}\right| < 1$), $\hat{\Omega}_{t+1}^\ell$ is scaled up relative to $\hat{\Omega}_{t+1}$. Conceptually, this means liquidity adjustment compensates for liquidity swings: if liquidity suddenly worsens, the adjusted model attributes a portion of the observed volatility to illiquidity and "smoothens" it out, yielding a lower covariance; if liquidity suddenly improves, the adjusted model attempts to extract underlying volatility masked by the steady condition, raising the covariance to a higher level. The smoothening effect makes the day-to-day conditional covariance more stable (less reactive to one-off liquidity shocks). This highlights the significance of the portfolio liquidity jump, that it quantifies the degree of liquidity adjustment in restoring the autoregressive continuity to the portfolio conditional covariance.

To verify the smoothening effect empirically, we compare the determinants (as a summary measure of size) of $\hat{\Omega}^{\ell}_{t+1}$ and $\hat{\Omega}_{t+1}$, $|\hat{\Omega}^{\ell}_{t+1}|$ and $|\hat{\Omega}_{t+1}|$ for both the cryptocurrency and stock portfolios. Based on the statistics of $|B^{\mathcal{P}\ell}_{r_t}|$ for the cryptocurrency portfolio (Panel A of Table 1, "liquidity jump $|B^{\mathcal{P}\ell}_{r_t}|$" column), we find that although its mean is higher than 1.0 (3.77), its median is less than 1.0 (0.70), and there are more days with a value less than 1.0 (1,191 or 53.12% of 2,242 days). As such, we hypothesize that that the determinant of the liquidity-adjusted $\hat{\Omega}^{\ell}_{t+1}$, $|\hat{\Omega}^{\ell}_{t+1}|$, increases from the determinant of $\hat{\Omega}_{t+1}$, $|\hat{\Omega}_{t+1}|$, for the cryptocurrency portfolio (alternative hypothesis: $|\hat{\Omega}_{t+1}| - |\hat{\Omega}^{\ell}_{t+1}| < 0$). Similarly, based on the statistics of $|B^{\mathcal{P}\ell}_{r_t}|$ for the stock portfolio (Panel B of Table 1, "liquidity jump $|B^{\mathcal{P}\ell}_{r_t}|$" column), we also hypothesize the same (alternative hypothesis: $|\hat{\Omega}_{t+1}| - |\hat{\Omega}^{\ell}_{t+1}| < 0$). We conduct a set of one-sided *t*-tests to compare $\hat{\Omega}^{\ell}_{t+1}$ and $\hat{\Omega}_{t+1}$, and present the results in Panel A of Table 2 for both the cryptocurrency and stock portfolios. We find that the alternative hypothesis is supported for the cryptocurrency portfolio, that $|\hat{\Omega}^{\ell}_{t+1}|$ increases from $|\hat{\Omega}_{t+1}|$ ($|\hat{\Omega}_{t+1}| - |\hat{\Omega}^{\ell}_{t+1}| < 0$) at the 5% significance level. We also find that the alternative hypothesis is supported for the stock portfolio, that $|\hat{\Omega}^{\ell}_{t+1}|$ increases from $|\hat{\Omega}_{t+1}|$ ($|\hat{\Omega}_{t+1}| - |\hat{\Omega}^{\ell}_{t+1}| < 0$) at the 10% significance level.

These results indicate that liquidity adjustment does indeed smoothen the conditional covariance for both portfolios, but the effect is much more pronounced for the cryptocurrency portfolio. That the significance level of the stock portfolio (10%) is lower than that of the cryptocurrency portfolio (5%) indicates that the regular return vector $Q_t$ (through constituent asset-level returns $r_{t_i}$'s) of the former already has an adequate amount of liquidity information implicitly priced in by the market, and therefore the liquidity adjustment does not add an extra amount of liquidity information to the return vector. Consequently, for the highly liquid stocks, although the liquidity-adjusted VECM-DCC/ADCC model does provide a better perspective on the portfolio-level conditional volatility than the regular model, the improvement is marginal. A more intuitive interpretation is that, for the cryptocurrency portfolio, unadjusted model leads the model to understate liquidity-driven volatility on some days. The adjusted model reduces this misspecification, resulting in a systematically "higher" covariance. For the stock portfolio, since liquidity is steady, both models give similar covariance, with only a slight difference.

In order to provide analytical explanation on the difference between the significant levels of the cryptocurrency portfolio (5%) and the stock portfolio (10%), we further investigate the impact of liquidity adjustment on correlation dynamics. We refer to Equations A2-3a and A2-3b in Appendix 2 for the DCC(1,1) and ADCC(1,1) specifications, in which the coefficients $a$, $b$, and $g$ reveal how correlation evolves over time. The "shock sensitive coefficient" $a$ captures the short-term responsiveness to new shocks, the "correlation persistence coefficient" $b$ reflects the persistence in correlation, and the "negative shock sensitive coefficient" $g$ (ADCC only) accounts for the asymmetric effect of negative shocks. We conduct a set of two-sided $t$-tests to study how liquidity adjustment affects the parameters, and present the results in Table 3.

In Panel A of Table 3, for the cryptocurrency portfolio, liquidity adjustment plays a critical role in refining these estimates. In the DCC model, liquidity adjustment significantly reduces $a$, $b$, and their sum $a + b$ at 1% significance level. These results suggest that, without liquidity adjustment, the DCC model overstates both the immediate responsiveness to shocks and the persistence of correlation dynamics, likely a reflection of the microstructure noise caused by high liquidity variability. In the ADCC model, while liquidity adjustment does not affect the symmetric shock parameter $a$ significantly ($t\_value = -0.07, p\_value = 0.94$), the persistence parameter $b$ and the asymmetry parameter $g$ both decrease at 1% significance level, as does the total dynamic component $a + b + g$. These findings indicate that liquidity adjustment help isolate genuine co-movements from distortions driven by high liquidity fluctuations, which underscores that liquidity adjustment is necessary in modeling the correlation dynamics of the cryptocurrency portfolio.

In contrast, while liquidity adjustment also impacts correlation dynamics in the stock portfolio, the effects are quantitatively different and less structurally necessary (Panel B of Table 3). For the DCC model, liquidity adjustment reduces the shock sensitivity parameter $a$ (1% significance) but increases the persistence parameter $b$ (1% significance), resulting in a net increase in $a + b$ (1% significance). This suggests that correlation dynamics become less reactive but more stable after accounting for liquidity. A similar pattern is observed in the ADCC model: liquidity adjustment reduces $a$ (1% significance) and $g$ (1% significance), and increases $b$ (1% significance), with the total dynamic component $a + b + g$ also increasing at 1% significance level. These adjustments indicate that in highly liquid stock markets, liquidity adjustment helps temper excessive reactivity while enhancing correlation persistence. However, it is not essential for avoiding model misspecification in the stock portfolios to the same degree as in the cryptocurrency portfolio.

Overall, the above findings demonstrate that liquidity adjustment plays a central role in ensuring the reliability of dynamic correlation models for asset classes like cryptocurrencies, where liquidity fluctuations introduce considerable short-term noise. For the cryptocurrency portfolio, failing to adjust for liquidity can result in overstated reactivity, misleading correlation persistence, and inflated asymmetry in estimated correlations. In contrast, the stock portfolio benefits from liquidity adjustment primarily in terms of improved stability, rather than correction of severe misspecification. These results emphasize the importance of tailoring econometric models to asset class characteristics, and highlight the broader implication that liquidity is not merely a microstructure detail but a fundamental determinant of reliable correlation modeling for assets with high liquidity variability, i.e., cryptocurrencies.

**5.4 Effect of Liquidity Adjustment on Posterior Covariance Estimate**

In addition, we establish the connection between the liquidity-adjusted Bayesian posterior covariance matrix and as follow (see Appendix 2 Subsection A2.3 for details):

$$\hat{\Sigma}^{TT}_{r^\ell_{t+1}} = B^{\mathcal{P}\ell^{-1}}_{\sigma_t} \left[ \Sigma^{TT}_{r_t} + \left[ (\tau \Sigma^{TT}_{r_t})^{-1} + \left( B^{\mathcal{P}\ell}_t \hat{\Omega}_{t+1} B^{\mathcal{P}\ell^H}_t \right)^{-1} \right]^{-1} \right] B^{\mathcal{P}\ell^{H^{-1}}}_{\sigma_t} \qquad (20)$$

$$\text{where: } B^{\mathcal{P}\ell}_t = B^{\mathcal{P}\ell}_{\sigma_t} B^{\mathcal{P}\ell^{-\frac{1}{2}}}_{r_t} \qquad (21)$$

In Equation 21, we establish a third matrix, the "portfolio liquidity composite matrix" $B^{\mathcal{P}\ell}_t$, which is the matrix product of portfolio liquidity diffusion matrix $B^{\mathcal{P}\ell}_{\sigma_t}$ and the inverse of the square root of portfolio liquidity jump matrix $B^{\mathcal{P}\ell}_{r_t}$, and is a scaling factor matrix for the regular conditional covariance $\hat{\Omega}_{t+1}$ in constructing the posterior covariance $\hat{\Sigma}^{TT}_{r^\ell_{t+1}}$. It is thus apparent that both the portfolio liquidity jump and diffusion have direct role in estimating the posterior portfolio covariance matrix. The portfolio liquidity matrix $B^{\mathcal{P}\ell}_t$ reflects that the two liquidity measure matrices work towards opposite directions in forming it. The descriptive statistics of the determinant of portfolio liquidity matrix $B^{\mathcal{P}\ell}_t$, $\left| B^{\mathcal{P}\ell}_t \right|$, is given in Table 1 "liquidity composite $\left| B^{\mathcal{P}\ell}_t \right|$" column (Panel A for the cryptocurrency portfolio, Panel B for the stock portfolio).

Equation 20 does not directly link the liquidity-adjusted posterior covariance matrix $\hat{\Sigma}^{TT}_{r^\ell_{t+1}}$ to its regular counterpart $\hat{\Sigma}^{TT}_{r_{t+1}}$. However, based on the statistics of $\left| B^{\mathcal{P}\ell}_{\sigma_t} \right|$ ("liquidity diffusion $\left| B^{\mathcal{P}\ell}_{\sigma_t} \right|$" column in Table 1 Panel A for cryptocurrency portfolio, Panel B for stock portfolio), Equation 20 demonstrates that when portfolio liquidity diffusion is high ($|B_{\sigma_t}^{\mathcal{P}\ell}| > 1$), the liquidity-adjusted posterior covariance $\hat{\Sigma}_{r_{t+1}^{\ell}}^{TT}$ is reduced from its regular counterpart $\hat{\Sigma}_{r_{t+1}}^{TT}$, and when it is low ($|B_{\sigma_t}^{\mathcal{P}\ell}| < 1$) the opposite is true. Conceptually this means that a high $|B_{\sigma_t}^{\mathcal{P}\ell}|$ indicates high intraday liquidity volatility, and more shrinkage is thus applied to $\hat{\Sigma}_{r_{t+1}^{\ell}}^{TT}$, rendering it smaller relative to $\hat{\Sigma}_{r_{t+1}}^{TT}$; and that a low $|B_{\sigma_t}^{\mathcal{P}\ell}|$ suggests steady intraday liquidity and the model trusts shrinks the volatility estimate less, resulting in a higher $\hat{\Sigma}_{r_{t+1}^{\ell}}^{TT}$ relative to $\hat{\Sigma}_{r_{t+1}}^{TT}$. In essence, liquidity adjustment tends to "smoothen" the posterior covariance when there is high liquidity variability. This signifies the importance of portfolio liquidity diffusion $B_{\sigma_t}^{\mathcal{P}\ell}$, that it quantifies the degree of liquidity adjustment in adjusting the portfolio-level autoregressive continuity to the posterior covariance.

We formally test the difference in determinants for the posterior covariance. Based on the statistics of $|B_{\sigma_t}^{\mathcal{P}\ell}|$ for the cryptocurrency portfolio (Panel A of Table 1, "liquidity jump $|B_{\sigma_t}^{\mathcal{P}\ell}|$" column), we find that there are more days with a value less than 1.0 (1,431 or 63.83% of 2,242 days). Therefore, we hypothesize that the determinant of $\hat{\Sigma}_{r_{t+1}^{\ell}}^{TT}$, $|\hat{\Sigma}_{r_{t+1}^{\ell}}^{TT}|$ is increases from the determinant of $\hat{\Sigma}_{r_{t+1}}^{TT}$, $|\hat{\Sigma}_{r_{t+1}}^{TT}|$ (alternative hypothesis: $|\hat{\Sigma}_{r_{t+1}}^{TT}| - |\hat{\Sigma}_{r_{t+1}^{\ell}}^{TT}| < 0$) for the cryptocurrency portfolio. With the same argument and based on the statistics of $|B_{\sigma_t}^{\mathcal{P}\ell}|$ for the stock portfolio (Panel B of Table 1, "liquidity jump $|B_{\sigma_t}^{\mathcal{P}\ell}|$" column), we also hypothesize same (alternative hypothesis: $|\hat{\Sigma}_{r_{t+1}}^{TT}| - |\hat{\Sigma}_{r_{t+1}^{\ell}}^{TT}| < 0$). We then conduct a set of one-sided $t$-tests to compare $|\hat{\Sigma}_{r_{t+1}}^{TT}|$ and $|\hat{\Sigma}_{r_{t+1}^{\ell}}^{TT}|$, and present the results in Panel B of Table 2. We find that the alternative hypothesis is supported for the cryptocurrency portfolio ($|\hat{\Sigma}_{r_{t+1}}^{TT}| - |\hat{\Sigma}_{r_{t+1}^{\ell}}^{TT}| < 0$) at the 10% significance level.

However, it is not supported for the stock portfolio based on its *p*-value at 0.16 for all DCC specifications, although the *t*-value is indeed negative ($\left|\hat{\Sigma}_{r_{t+1}}^{TT}\right| - \left|\hat{\Sigma}_{r_{t+1}^{\ell}}^{TT}\right| < 0$ still holds). The lack of significance for stocks is hardly surprising given the near-zero $\left|B_{\sigma_t}^{\mathcal{P}\ell}\right|$ (Panel B of Table 1, "liquidity diffusion $\left|B_{\sigma_t}^{\mathcal{P}\ell}\right|$" column), as there is almost no intraday liquidity noise to correct for, thus the Bayesian posterior largely stays the same with or without the liquidity info.

The interpretation of the one-sided *t*-test results is analogous to the conditional covariance case. That the test does not yield a statistically significant support to the alternative hypothesis for the stock portfolio indicates that the regular posterior covariance $\hat{\Sigma}_{r_{t+1}}^{TT}$ of the stock portfolio already has an adequate amount of liquidity information implicitly priced in by the market, much like in the case of the regular return vector $Q_t$, and therefore the liquidity adjustment does not add a statistically significant amount of extra liquidity information to posterior covariance. Consequently, for US stocks, the liquidity-adjusted Bayesian model does not provide a better perspective on the portfolio-level posterior volatility than the regular model.

As such, we provide empirical evidence that, for the cryptocurrency portfolio, liquidity adjustment smoothens the conditional covariance in the autoregressive VECM-DCC/ADCC model through the scaling effect of portfolio liquidity jump ($B_{r_t}^{\mathcal{P}\ell}$), and the posterior covariance in the Bayesian model through the scaling effect of portfolio liquidity diffusion ($B_{\sigma_t}^{\mathcal{P}\ell}$). For the stock portfolio, however, liquidity adjustment only smoothens the conditional covariance in the autoregressive VECM-DCC/ADCC model, albeit in a more marginal manner, but has no statistically significant impact on the posterior covariance in the Bayesian model.

The results supports our core proposition: if the return and volatility of assets with high liquidity variability are properly adjusted by liquidity, the magnitude of liquidity fluctuation is reduced (scaled down by portfolio liquidity jump) and the DCC/ADCC model produces smoother portfolio conditional covariance, and the volatility of liquidity fluctuation is also reduced (scaled down by portfolio liquidity diffusion) and the Bayesian model produces less irrational expected increment in the portfolio posterior covariance. In Section 6, we provide additional empirical support to our proposition with a set of comparative tests between the TMV and LAMV portfolios for both cryptocurrency and stock portfolios.

## 6. Empirical Tests with Mean-Variance Portfolio Optimization

In this section we provide further empirical evidence that the liquidity-adjusted models of Section 5 offer better predictability on posterior portfolio covariance matrix than their traditional counterparts. We compare a series of mean-variance optimized portfolios to evaluate whether using liquidity-adjusted inputs (LAMV portfolios) yields better out-of-sample performance than traditional approaches (TMV portfolios). We construct six portfolios for both asset classes (cryptocurrency and stock), each with progressively more sophisticated covariance inputs.

### 6.1 Standard MV Portfolios

We first construct two MV portfolios: traditional and liquidity-adjusted. The standard daily-optimized MV in a time-series construct analytically expressed as the following quadratic programming problem with constraints:

$$\max_{W_t} \left( \bar{\mu}_t W_t - \frac{\lambda_t}{2} W_t^H \bar{\Sigma}_t W_t \right); H \text{ is Transpose} \quad (21)$$

subject to:
$w_t^{r_f} + \sum_i^N w_t^i = 1$; $i = $ list of assets; $N = $ number of assets; $r_f$ is risk-free asset
$w_t^i, w_t^{r_f} \geq 0$ (long − only)
$w_t^{r_f} \leq 1$
$w_t^i \leq \frac{3}{N}$ (equal weight)

*where:*

$$\lambda_t = \frac{r^P_{t_{mkt}} - r^{rf}_t}{\sigma^2_{t\,mkt}} = \frac{r^P_{t_{mkt}}}{\sigma^2_{t\,mkt}}$$

$r^P_{t_{mkt}}$ is the return of the market or equilibrium portfolio on day t, $\sigma^2_{t\,mkt}$ is its variance of the rolling window; $r^{rf}_t$ is the return of risk-free asset, regarded as being 0%.

In the standard MV construct of Equation 21, $\bar{\mu}_t$ is the portfolio mean return vector over a window ending on day $t$, and $\bar{\Sigma}_t$ is the covariance matrix of daily returns of the constituent assets in that rolling window (note: it is not the daily minute-level covariance matrix, $\Sigma^{TT}_t$, see the rest of this section). Both $\bar{\mu}_t$ and $\bar{\Sigma}_t$ are realized and derived from available information up to day $t$. In addition, $W_t$ is the portfolio (column) weight vector to be optimized for day $t$. The daily MV portfolios are:

1. Portfolio 1: standard TMV portfolio; $\bar{\mu}_t$ is the mean vector of $r_t$'s over the rolling window ending on day $t$, or $\bar{\mu}_{r_t}$; $\bar{\Sigma}_t$ is the covariance matrix of $r_t$'s for the rolling window, or $\bar{\Sigma}_{r_t}$.

2. Portfolio 2: standard LAMV portfolio; $\bar{\mu}_t$ is the mean vector of $r^\ell_t$'s over the rolling window ending on day $t$, or $\bar{\mu}_{r^\ell_t}$; $\bar{\Sigma}_t$ is the covariance matrix of $r^\ell_t$'s for the rolling window, or $\bar{\Sigma}_{r^\ell_t}$.

## 6.2 Intraday Covariance Matrix MV Portfolios

To demonstrate the utilities of the intraday covariance matrix in portfolio performance, built upon Portfolios 1 and 2, we further construct two MV portfolios with intraday covariance matrix by rewriting Equation 21 to retain $\bar{\mu}_t$ and to replace $\bar{\Sigma}_t$ by the intraday covariance matrix on day t, $\Sigma^{TT}_t$. The portfolios are constructed as:

$$\max_{W_t} \left( \bar{\mu}_t W_t - \frac{\lambda_t}{2} W^H_t \Sigma^{TT}_t W_t \right) \tag{22}$$

All the constraints for Equation 22 are the same as those for Equation 21. The intraday covariance matrix MV portfolios are:

3. Portfolio 3: intraday TMV portfolio; $\bar{\mu}_t$ is the mean vector of $r_t$'s over the rolling window ending on day $t$, $\bar{\mu}_{r_t}$; $\Sigma_t^{TT}$ is the regular minute-level (intraday) covariance matrix, $\Sigma_{r_t}^{TT}$.

4. Portfolio 4: intraday LAMV portfolio; $\bar{\mu}_t$ is the mean vector of $r_t^\ell$'s over the rolling window ending, $\bar{\mu}_{r_t^\ell}$; $\Sigma_t^{TT}$ is the liquidity-adjusted minute-level (intraday) covariance matrix, $\Sigma_{r_t^\ell}^{TT}$.

## 6.3 VECM-DCC/ADCC-Bayesian-enhanced MV Portfolios

To demonstrate the utilities of the liquidity-adjusted VECM-DCC/ADCC-Bayesian framework of Section 5, we further construct two MV portfolios with VECM-DCC/ADCC-Bayesian enhancement. We rewrite Equation 21 to replace $\bar{\Sigma}_t$ by the forecasted posterior covariance matrix on day $t+1$, $\hat{\Sigma}_{t+1}^{TT}$. The portfolios are constructed as:

$$\max_{W_t} \left( \bar{\mu}_t W_t - \frac{\lambda_t}{2} W_t^H \hat{\Sigma}_{t+1}^{TT} W_t \right) \quad (23)$$

All the constraints for Equation 23 are the same as those for Equation 21. The VECM/VAR-DCC/ADCC-Bayesian-enhanced MV portfolios are:

5. Portfolio 5: enhanced TMV portfolio; $\bar{\mu}_t$ is the mean vector of $r_t$'s over the rolling window ending on day $t$, or $\bar{\mu}_{r_t}$; $\hat{\Sigma}_{t+1}^{TT}$ is the forecasted daily regular minute-level (intraday) covariance matrix for day $t+1$, $\hat{\Sigma}_{r_{t+1}}^{TT}$.

6. Portfolio 6: enhanced LAMV portfolio; $\bar{\mu}_t$ is the mean vector of $r_t^\ell$'s over the rolling window ending on day $t$, or $\bar{\mu}_{r_t^\ell}$; $\hat{\Sigma}_{t+1}^{TT}$ is the forecasted daily liquidity-adjusted minute-level (intraday) covariance matrix, $\hat{\Sigma}_{r_{t+1}^\ell}^{TT}$.

## 6.4 Performance Comparisons of Cryptocurrency Portfolios

We use the annualized Sharpe Ratio ($SR_a$) to compare the performance between portfolios:

$$SR_a = \frac{r_a^P - r_a^{rf}}{\sigma_a^P} = \frac{r_a^P}{\sigma_a^P} \quad (23)$$

Where:
1. $r_a^P, \sigma_a^P$ are the annualized realized regular daily portfolio return and standard deviation.
2. $r_a^{rf}$ us the annualized realized daily returns for the risk-free asset.

In Table 4 we capture the descriptive statistics of daily portfolio return, volatility and Sharpe Ratio (Panel A) of the 6 MV portfolios (Portfolios 1 to 6), which are arranged as such: the TMV portfolios with incremental forecast enhancement are listed on the left (Portfolios 1, 3, 5), while their corresponding LAMV portfolios are shown on the right (Portfolios 2, 4, 6). That way, it is easier to observe the improvement after applying each type of enhancement methodology vertically within the TMV and LAMV, while at the same time conveniently compare the differences between the TMV and LAMV after applying each specific incremental enhancement methodology horizontally.

**6.4.1 TMV Portfolios and Limitations of Risk Modeling without Liquidity Adjustment**

Panel A of Table 4 summarizes the performance of all 6 cryptocurrency portfolios. The TMV portfolios exhibit a performance deterioration in general. Among the TMV portfolios, the standard TMV portfolio (Portfolio 1) has an annualized Sharpe Ratio ($SR_a$) of 0.83. With the intraday covariance matrix $\Sigma_t^{TT}$ replacing the interday (rolling window) covariance matrix $\bar{\Sigma}_t$, the intraday TMV Portfolio 3 has a large drop of $SR_a$ at 0.52. With the forecasted intraday covariance matrix $\hat{\Sigma}_{r_{t+1}}^{TT}$ replacing the intraday covariance matrix $\Sigma_t^{TT}$, the full VCEM-DCC/ADCC-Bayesian-enhanced TMV Portfolio 5 has a noticeably improved $SR_a$ of 0.76 from Portfolio 3 (0.52), which is still much lower than that of Portfolio 1 (0.83).

These empirical results show that increased sophistication in covariance estimation does not necessarily lead to better performance when liquidity is not accounted. Despite being the most basic, Portfolio 1 achieves the highest $SR_a$ (0.83), outperforming both Portfolio 3 (0.52) and Portfolio 5 (0.76). Portfolio 3 exhibits the weakest performance, with significantly lower return and comparable volatility, suggesting that static intraday covariance matrices may overreact to transient volatility and produce overly conservative allocations. Portfolio 5 performs better than Portfolio 3, benefiting from dynamic forecast adjustments, but still underperforms the simplest model, Portfolio 1. This pattern suggests a fundamental misalignment between the return and risk modeling components in the TMV framework. Advanced risk models like VECMDCC/ADCC-Bayesian updates may improve volatility estimates, but without corresponding adjustments to reflect liquidity conditions, the MV optimizer may overweight high-risk and illiquid assets. The result is inefficient capital allocation and diminished portfolio efficiency. In summary, the comparisons among TMV portfolios reveal that, in the absence of liquidity-aware modeling, improvements in risk estimation alone are insufficient to enhance portfolio performance.

**6.4.2 LAMV Portfolios and Values of Dynamic Risk Estimation with Liquidity Adjustment**

Unlike the case of TMV portfolios, the LAMV portfolios demonstrate a clear incremental improvement with each enhancement. The standard LAMV Portfolio 2 provides a liquidity-aware baseline, achieving moderate volatility control but relatively modest returns and a $SR_a$ of 0.67. The intraday LAMV Portfolio 4 exhibits a significant improvement across all metrics, achieving a $SR_a$ of 0.94. The realized intraday covariance matrix captures high-frequency dynamics more accurately, supporting better reallocation of capital in response to short-term market shifts. By combining liquidity-adjusted return with forecasted covariance matrix, the enhanced LAMV Portfolio 6 shows the best overall performance with the highest $SR_a$ at 1.04. The improved performance reflects the ability of VECM-DCC/ADCC-Bayesian model to incorporate both recent volatility behavior while preserving the liquidity structure of returns. This allows the MV optimizer to balance exposure to risk and liquidity more effectively than static models. The results clearly indicate that when return and volatility reflect liquidity conditions, improvements in covariance estimation produce meaningful performance gains. Taken together, the performance of the LAMV portfolios demonstrate that sophisticated risk models can fully realize their potential when aligned with return and volatility that incorporate liquidity conditions and effects. The combination of liquidity-adjusted return and volatility and advanced dynamic risk modeling results in superior portfolio efficiency.

**6.4.3 TMV vs. LAMV: The Importance of Return-Risk Alignment**

A comparison between the TMV and LAMV portfolio sets reveals an important structural insight: advanced risk models can only enhance portfolio performance when the return and volatility components are adjusted to reflect liquidity conditions. This is most clearly illustrated by comparing the of corresponding portfolios under each framework. For the rolling-window specification, the standard TMV Portfolio 1 actually outperforms the standard LAMV Portfolio 2 in $SR_a$ (0.83 vs. 0.67), suggesting that a liquidity-adjusted return vector may lead to under allocation to high-return assets when paired with static risk models.

However, the advantage of the LAMV framework becomes pronounced as the sophistication of the risk model increases. With realized intraday covariance matrices, $SR_a$ improves from 0.52 in intraday TMV Portfolio 3 to 0.94 in intraday LAMV Portfolio 4. Similarly, when the full VECM-DCC/ADCC-Bayesian forecasted covariance matrix is used, $SR_a$ rises from 0.76 (TMV Portfolio 5) to 1.04 (LAMV Portfolio 6). These improvements indicate that liquidity-aware return and volatility help correct the overly conservative tendency of intraday models and align capital allocation with actual liquidity conditions. The results suggest that dynamic covariance estimators such as DCC-ADCC and Bayesian models are most effective when used within a framework that also accounts for liquidity-adjusted return and volatility. Without this alignment, even the best risk forecasts may guide the MV optimizer toward mispriced or illiquid assets, reducing the effectiveness of portfolio allocation.

In conclusion, the comparisons between the TMV and LAMV portfolios highlights a key argument of this paper: for portfolios of assets exposed to extreme liquidity variability, improvements in risk modeling only translate into performance gains when return and volatility expectations are also conditioned on liquidity. It is the joint refinement of both liquidity adjustment and advanced risk modeling that leads to robust and effective portfolio strategies for assets with extreme liquidity variability.

### 6.5 Performance Comparisons of Stock Portfolios

To assess whether the findings from cryptocurrency portfolios extend to more traditional asset portfolios, we analyze the performance of analogous TMV and LAMV portfolios of U.S. stocks. The construction of these portfolios is identical in methodology but applied to a markedly different asset class, characterized by higher level of liquidity stability. The TMV portfolios using regular return show generally strong performance. The standard TMV Portfolio 1 has a $SR_a$ of 1.20. Interestingly, the intraday TMV Portfolio 3 achieves a slightly higher $SR_a$ of 1.27, while fully enhanced TMV Portfolio 5 performs slightly below that at 1.21. This suggests that, unlike in crypto markets, more advanced covariance modeling can offer incremental benefits even without liquidity-adjusted returns. For the LAMV portfolios, the standard LAMV Portfolio 2 slightly underperforms its TMV counterpart in $SR_a$ (1.18 vs. 1.20), primarily due to reduced return. However, as in the cryptocurrency analysis, the LAMV portfolios benefit significantly from enhanced risk modeling, as $SR_a$ increases to 1.30 with the intraday LAMV Portfolio 4 and reaches a peak of 1.31 with the fully enhanced LAMV Portfolio 6.

These results reaffirm the central insight from the previous sections: liquidity-adjusted return and volatility enhance the value of advanced risk modeling. However, for US stocks, of which the liquidity fluctuation is less extreme than that of cryptocurrencies, the improvements are more modest in absolute terms. The gains are visible particularly in liquidity-adjusted return rather than regular return, with volatility declining slightly across all LAMV portfolios. Thus, while TMV performs relatively well in traditional markets, the LAMV framework still offers a narrow edge, especially when paired with dynamic risk forecasts. The improvement is smaller than in high-volatility, high-friction asset classes, but the direction and structure of the gains are consistent.

**6.6 Cross Asset Comparative Analysis: Cryptocurrency vs. Stock Portfolio Dynamics**

When comparing portfolio performance across asset classes, a few patterns emerge. For cryptocurrency portfolios, the LAMV portfolios produce substantial incremental improvements in $SR_a$ only when combined with fully-enhanced covariance estimators. In contrast, for stock portfolios, while the LAMV portfolios still offers consistent incremental enhancements in performance, their advantage over the TMV portfolios is modest. The comparisons between asset classes underscore a broader theme: the greater the liquidity variability and execution friction in an asset class, the more critical it becomes to align both the return and risk components of the portfolio optimization model with liquidity conditions. The LAMV framework augmented with VECM-DCC/ADCC-Bayesian covariance estimation offers a robust and adaptable solution that scales across markets with varying microstructural characteristics.

**7. Conclusions**

Cryptocurrencies have emerged as an increasingly important asset class in the modern financial system, fueled by rapid institutional adoption, global accessibility, and distinctive return-generating characteristics. In contrast to traditional assets such as US stocks, cryptocurrencies operate in fragmented, continuously trading (24/7) markets that exhibit pronounced liquidity variability. This complexity poses fundamental challenges to conventional modeling frameworks at both the asset and portfolio levels. The extreme variability, discontinuity, and episodic nature of liquidity in these markets violate the core assumptions of standard return and volatility models, which assume normal and stationary dynamics. Without explicit liquidity adjustment, these models struggle to capture the time-varying risk structure and cross-asset co-movement inherent in cryptocurrency trading. Our study highlights the necessity of a liquidity-adjusted modeling framework to improve risk estimation and portfolio construction under such conditions.

In this paper, we propose and empirically validate a liquidity-sensitive modeling framework that enhances the predictability of multivariate return and volatility, especially for portfolios exposed to extreme liquidity variability. We begin by constructing liquidity-adjusted return and volatility measures that reflect real-time market frictions and better approximate the underlying return-generating process. From these inputs, we derive two novel portfolio-level liquidity measures: portfolio liquidity jump and portfolio liquidity diffusion, which quantify the magnitude and volatility of liquidity variation across assets, respectively. We thoroughly examine the distribution of both portfolio liquidity metrics for two asset classes, cryptocurrency and US stock. These measures offer dynamic insights into liquidity conditions and serve as key indicators for enhancing econometric modeling.

We then develop a unified VECM-DCC/ADCC-Bayesian framework, applying it to both regular and liquidity-adjusted return and volatility series for the two asset classes with contrasting liquidity profiles. Our empirical results show that liquidity-adjusted inputs significantly stabilize correlation dynamics and improve forecasting accuracy, especially in the case of the cryptocurrency portfolio, of which the liquidity variation is both extreme and persistent. When integrated into this econometric structure, liquidity-adjusted return and volatility restore the effectiveness of risk modeling for assets under stressed liquidity conditions. In essence, the framework adapts traditional autoregressive models to the liquidity realities of the market, yielding more reliable estimates of evolving portfolio risk.

Within a classical mean-variance optimization setting, we demonstrate that portfolios optimized using our liquidity-adjusted model (LAMV portfolios) consistently outperform their traditional counterparts (TMV portfolios) across both asset classes. The performance gains are particularly pronounced for cryptocurrencies, where the absence of liquidity adjustment in the TMV portfolios leads to significant model misspecification and suboptimal allocations, while the LAMV portfolios estimate the risk level appropriately, resulting in substantially better Sharpe Ratios. These findings underscore that, for assets with high liquidity variability, liquidity adjustment is not only beneficial but essential for robust multivariate risk modeling and effective portfolio design. Even for US stocks, of which the liquidity is relatively stable, the LAMV portfolios deliver modest yet meaningful improvements in Sharpe Ratios over TMV portfolios.

A consistent insight from our analysis is that advanced risk modeling techniques (in our study, DCC/ADCC for time-varying conditional covariance and Bayesian for posterior covariance) only reach their full potential when built upon liquidity-adjusted return and volatility. Without liquidity adjustment, enhanced modeling in risk estimation does not reliably translate into improved portfolio performance. Particularly for asset classes with volatile liquidity, ignoring liquidity can lead sophisticated models to draw the wrong conclusions (for example, misjudging when volatility spikes as ephemeral when they are structural). In contrast, incorporating liquidity conditions allows these models to align their outputs with investable reality.

Overall, our findings affirm that incorporating portfolio-level liquidity dynamics into multivariate time-series modeling and portfolio optimization produces more robust and effective investment strategies. The liquidity-adjusted framework we present is generalizable and can be extended to other asset classes facing liquidity challenges, such as corporate bonds or emerging market assets. It adds a crucial layer of realism to risk modeling, ensuring that both expected return and risk are evaluated in the context of prevailing liquidity conditions.

The proposed framework has implications for all market participants. For practitioners, accounting for liquidity is not just a precaution but can be essential for unlocking the full value of advanced portfolio models. For regulators and policymakers, our results highlight the systemic importance of liquidity: models that ignore it may underestimate risks in assets with high liquidity variability, whereas those that include it can better foresee stress scenarios. Academic researchers may apply our liquidity-adjusted framework to assess trading strategies, develop models for tail-risk measures, or examine optimal execution strategies informed by the proposed liquidity metrics.

In conclusion, this paper proposes a comprehensive framework that integrates liquidity conditions into multivariate volatility modeling and portfolio optimization. The evidence strongly suggests that doing so materially improves outcomes in environments with extreme liquidity variability, and yields a more reliable modeling paradigm. We contribute a step forward in bridging the gap between theoretical models and the practical reality of investing in liquidity-volatile asset classes, ultimately facilitating better risk management and asset allocation decisions.

# Appendix 1 – Deriving the Portfolio Liquidity Volatility Beta Matrix $B^{\mathcal{P}\ell}_{\sigma_t}$

Deng (2024) proposes a Conditional Singular Value Decomposition (Conditional SVD) in the form of $A_{\{mn\}} = H_{\{mk\}} B_{\{kl\}} M^*_{\{ln\}}$ for given general matrices $A_{\{mn\}}$ and $B_{\{kl\}}$, and provides a special case, that when $m = n = k = l$, a reduced conditional SVD of the following exists:

$$A = HBH^*; \quad \text{where: } A, B \in \mathbb{C}^{n \times n} \tag{A1-1}$$

$A$ and $B$ have the SVD decompositions as:

$$A = U_A \Sigma_A U_A^*; B = U_B \Sigma_B U_B^* \tag{A1-2a, 2b}$$

where:
$U's$ are square complex unitary matrices; $\Sigma's$ are rectangular diagonal matrices with non-negative real numbers.

And there exists a decomposition between $\Sigma_A$ and $\Sigma_B$ as:

$$\Sigma_A = R \Sigma_B R^* \Rightarrow \Sigma_A = RR^* \Sigma_B = RR \Sigma_B \Rightarrow R = (\Sigma_A \Sigma_B^{-1})^{\frac{1}{2}} \tag{A1-3}$$

where: $R$ is a diagonal matrix with real numbers

By substituting $\Sigma_A$ in Equation A1-2a with the RHS of Equation A1-3 we get:

$$A = U_A (R \Sigma_B R^*) U_A^* \Rightarrow A = (U_A R) \Sigma_B (U_A R)^* \tag{A1-4a}$$

Also, substitute $B$ in Equation A1-1 with the RHS of Equation A1-2b:

$$A = H (U_B \Sigma_B U_B^*) H^* \Rightarrow A = (H U_B) \Sigma_B (H U_B)^* \tag{A1-4b}$$

By comparing Equation A1-4a and Equation A1-4b we get:

$$U_A R = H U_B \Rightarrow H = U_A R U_B^* \Rightarrow H = U_A (\Sigma_A \Sigma_B^{-1})^{\frac{1}{2}} U_B^* \tag{A1-5}$$

Equation A1-5 solves $H$ in proposition $A = HBH^*$, with $HH^*$ being a symmetric matrix.

Equation 7 in Subsection 5.2 is a special case of the special case in Deng (2024), in which $A$, $B$ are symmetric square matrices with non-negative elements, and non-zero values on the diagonal:

Let $\Sigma^{TT}_{r_t} \equiv A$, $\Sigma^{TT}_{r_t^\ell} \equiv B$ and $B^{\mathcal{P}\ell}_{\sigma_t} \equiv H$, we get:

$$\Sigma^{TT}_{r_t} = B^{\mathcal{P}\ell}_{\sigma_t} \Sigma^{TT}_{r_t^\ell} B^{\mathcal{P}\ell H}_{\sigma_t}; \quad \text{where: } \Sigma^{TT}_{r_t} = U_A \Sigma_A U_A^H; \Sigma^{TT}_{r_t^\ell} = U_B \Sigma_B U_B^H; \Sigma_A = R \Sigma_B R^H \tag{A1-6a}$$

$$B^{\mathcal{P}\ell}_{\sigma_t} = U_A R U_B^{-1} \tag{A1-6b}$$

# Appendix 2 – Liquidity-Adjusted VECM-DCC/ADCC-Bayesian Model

## A2.1 Regular VECM/VAR-DCC/ADCC-Bayesian Model

The full expression of regular VECM/VAR($p$) specification given by Equation 8 is:

$$Q_t = \sum_{i=1}^{p} \Phi_i Q_{t-i} + E_t \tag{A2-1}$$

$$\Delta Q_t = \Gamma Q_{t-1} + \sum_{i=1}^{p-1} \Phi_i^* \Delta Q_{t-i} + E_t$$

$$\Phi_i^* = -\sum_{j=i+1}^{p} \Phi_i, i = 1, \dots, p-1$$

$$\Gamma = -\left(I - \sum_{i=1}^{p} \Phi_i\right) = -\Phi(1)$$

*Where $Q_t$ is the portfolio return vector, $E_t$ is the residual vector, and $\Phi_i$ is the coefficient matrix for VAR lag i.*

The regular VECM/VAR($p$) specification of Equation A2-1 produces a one-period ($t+1$) forecasted return vector, $\hat{Q}_{t+1}$, of which the residual vector, $\hat{E}_{t+1}$, is given by Equation 9:

$$\hat{E}_{t+1} = \hat{Q}_{t+1} - Q_{t+1} \tag{A2-2}$$

*Where $Q_{t+1}$ is the actual observed return vector (out-of-sample observation) at time t+1.*

We then apply a DCC(1,1) specification to estimate the time-varying conditional covariance in the residual error vector $\hat{E}_{t+1}$. The DCC(1,1) specification is given as (Deng, 2018, Equation 5, with modifications on symbols):

$$\hat{E}_{t+1} | \Psi_t \sim N\left(0, \hat{\Omega}_{t+1} = \hat{H}_{t+1} \hat{P}_{t+1} \hat{H}_{t+1}\right) \tag{A2-3a}$$

$$\hat{H}_{t+1}^2 = H_0^2 + K E_t E_t^H + \Lambda H_t^2$$

$$\hat{P}_{t+1} = \hat{O}_{t+1}^* \hat{O}_{t+1} \hat{O}_{t+1}^*$$

$$\hat{O}_{t+1} = (1 - a - b)\bar{O} + a \Xi_t \Xi_t^H + b O_t$$

$$\Xi_t = H_t^{-1} E_t$$

$$a + b < 1$$

*Where:*
1) *$\hat{E}_{t+1}$ is the conditional residual vector from the VECM/VAR stage;*
2) *$\hat{\Omega}_{t+1}$ is the conditional covariance matrix of $\hat{E}_{t+1}$;*
3) *$\hat{H}_{t+1}$ is the normalization matrix for $\hat{P}_{t+1}$;*
4) *K and Λ are diagonal coefficient matrices for $H_t$;*
5) *$\hat{P}_{t+1}$ is the conditional correlation matrix of $\hat{E}_{t+1}$;*
6) *$\hat{O}_{t+1}$ and $\hat{O}_{t+1}^*$ are estimator matrices for $\hat{P}_{t+1}$;*
7) *$\bar{O}$ is the unconditional correlation matrix of $E_t$;*
8) *$O_t$ is the dynamic correlation matrix of $E_t$*
9) *$\Xi_t$ is the standardized residual vector of $E_t$.*

In order to accommodate asymmetries among conditional covariance and structural break induced conditional correlation increase, we also apply an ADCC(1,1) specification to $\hat{E}_{t+1}$. The ADCC(1,1) can be regarded as the multivariate variation of EGARCH(1,1) in that coefficient $g$ reflects sign impact (Deng, 2018, Equation 6, with modifications on symbols):

$$\hat{E}_{t+1}|\Psi_t \sim N(0, \hat{\Omega}_{t+1} = \hat{H}_{t+1}\hat{P}_{t+1}\hat{H}_{t+1}) \qquad (A2\text{-}3b)$$

$$\hat{H}_{t+1}^2 = H_0^2 + KE_t E_t^H + \Lambda H_t^2$$

$$\hat{P}_{t+1} = \hat{O}_{t+1}^* \hat{O}_{t+1} \hat{O}_{t+1}^*$$

$$\hat{O}_{t+1} = (1 - a - b)\bar{O} - g\bar{N} + a\Xi_t \Xi_t^H + bO_t + gN_t N_t^H$$

$$\Xi_t = H_t^{-1} E_t$$

$$N_t = I[\xi_{i,t} < 0] \circ \Xi_t$$

$$a + b + g < 1$$

Where:
1) $N_t$ augments the asymmetric effect of the negative elements $\xi_{i,t} < 0$ in $\Xi_t$;
2) the matrix operator "∘" is the Hadamard product of two identically sized matrices/vectors, computed simply by element-wise multiplication;
3) all other parameters are defined the same way as in Equation A4.

For each rolling window, we fit both DCC(1,1) and ADCC(1,1) on $\hat{E}_{t+1}$, and choose either DCC(1,1) or ADCC(1,1) with a higher log-likelihood to produce the conditional covariance matrix, $\hat{\Omega}_{t+1}$. With the forecasted $\hat{\Omega}_{t+1}$ we further estimate the posterior (forecasted) daily covariance matrix for day $t+1$, $\hat{\Sigma}_{r_{t+1}}^{TT}$, which is analytically expressed as (Deng, 2018, Equation 8, with modifications on symbols):

$$\hat{\Sigma}_{r_{t+1}}^{TT} = \Sigma_{r_t}^{TT} + \widehat{M}_{t+1}^{-1} \qquad (A2\text{-}4a)$$

$$\widehat{M}_{t+1}^{-1} = \left[\left(\tau \Sigma_{r_t}^{TT}\right)^{-1} + \widehat{Pm}_{t+1}^H \hat{\Omega}_{t+1}^{-1} \widehat{Pm}_{t+1}\right]^{-1} = \left[\left(\tau \Sigma_{r_t}^{TT}\right)^{-1} + I_{N \times N} \hat{\Omega}_{t+1}^{-1} I_{N \times N}\right]^{-1} = \left[\left(\tau \Sigma_{r_t}^{TT}\right)^{-1} + \hat{\Omega}_{t+1}^{-1}\right]^{-1} \qquad (A2\text{-}4b)$$

where:
1) $\widehat{M}_{t+1}^{-1}$ is the adjustment to the covariance matrix at time $t$ for the next time period $t+1$;
2) $\widehat{Pm}_{t+1}$ is the 1-period forward estimated weight matrix representing the investor's views and companion of $\hat{Q}_{t+1}$, thus it is a $N \times N$ matrix. Since $\hat{Q}_{t+1}$ is "absolute," as it is forecasted in an objective fashion, $\widehat{Pm}_{t+1}$ is an identity matrix of order $N$, $I_{N \times N}$, $N$ is the number of assets in the portfolio;
3) $\tau$ is the "confidence" parameter for the forecasted values. It is typically between 0.01 and 10, and we choose a value of 1.0 through experimentation.

We thus consolidate Equations A2-4a and A2-4b into Equation A2-4 and estimate the posterior regular daily regular covariance matrix for day $t+1$ as:

$$\hat{\Sigma}_{rt+1}^{TT} = \Sigma_{rt}^{TT} + \left[(\tau\Sigma_{rt}^{TT})^{-1} + \hat{\Omega}_{t+1}^{-1}\right]^{-1} \quad (A2\text{-}4)$$

Finally, we consolidate Equations A2-1, A2-2, A2-3a/3b and A2-4b of the regular VECM-DCC/ADCC-Bayesian model at extreme liquidity as:

$$Q_t = \sum_{i=1}^{p} \Phi_i Q_{t-i} + E_t \quad (A2\text{-}1)$$

$$\hat{E}_{t+1} = \hat{Q}_{t+1} - Q_{t+1} \quad (A2\text{-}2)$$

$$\hat{E}_{t+1}|\Psi_t \sim N(0, \hat{\Omega}_{t+1}) \quad (A2\text{-}3)$$

$$\hat{\Sigma}_{rt+1}^{TT} = \Sigma_{rt}^{TT} + \left[(\tau\Sigma_{rt}^{TT})^{-1} + \hat{\Omega}_{t+1}^{-1}\right]^{-1} \quad (A2\text{-}4)$$

Where:
1) $Q_t$ is the portfolio return vector, $E_t$ is the residual vector, and $\Phi_i$ is the coefficient matrix for VAR lag I,
2) $\hat{Q}_{t+1}$ is the forecasted portfolio return vector (out-of-sample) at time t+1,
3) $Q_{t+1}$ is the actual observed return vector (out-of-sample) at time t+1,
4) $\hat{E}_{t+1}$ is the conditional residual vector (out-of-sample) at time t_1, from the VECM/VAR stage,
5) $\hat{\Omega}_{t+1}$ is the conditional covariance matrix of $\hat{E}_{t+1}$ in the rolling window.

We then transfer Equations A2-1 to A2-4 back to Subsection 5.1 as Equations 8-11.

## A2.2 Liquidity-Adjusted VECM/VAR-DCC/ADCC-Bayesian Model

Equation 4 gives the connection between the daily regular and liquidity-adjusted return vectors, with the subscript substitution we have:

$$Q_t = B_{rt}^{\mathcal{P}\ell} Q_t^\ell \Rightarrow Q_t^\ell = B_{rt}^{\mathcal{P}\ell^{-1}} Q_t \quad (A2\text{-}5)$$

By substituting $Q_t$ and $Q_{t-i}$ in Equation A2-1 by the RHS of Equation A2-5 we get:

$$Q_t = \sum_{i=1}^{p} \Phi_i Q_{t-i} + E_t$$

$$\Rightarrow B_{rt}^{\mathcal{P}\ell} Q_t^\ell = \sum_{i=1}^{p} \Phi_i B_{rt-i}^{\mathcal{P}\ell} Q_{t-i}^\ell + E_t$$

$$\Rightarrow Q_t^\ell = \sum_{i=1}^{p} B_{rt}^{\mathcal{P}\ell^{-1}} \Phi_i B_{rt-i}^{\mathcal{P}\ell} Q_{t-i}^\ell + B_{rt}^{\mathcal{P}\ell^{-1}} E_t$$

$$\Rightarrow Q_t^\ell = \sum_{i=1}^{p} \Phi_i^\ell Q_{t-i}^\ell + E_t^\ell \quad (A2\text{-}6a)$$

$$\text{where: } \Phi_i^\ell = B_{rt}^{\mathcal{P}\ell^{-1}} \Phi_i B_{rt-i}^{\mathcal{P}\ell}; \; E_t^\ell = B_{rt}^{\mathcal{P}\ell^{-1}} E_t \quad (A2\text{-}6b)$$

The expected error vector $\hat{E}_{t+1}^\ell$ is given as:

$$\hat{E}_{t+1}^\ell = \hat{Q}_{t+1}^\ell - Q_{t+1}^\ell \quad (A2\text{-}7)$$

The conditional covariance matrix $\hat{\Omega}_{t+1}$ of the expected error vector $\hat{E}_{t+1}$ is given by Equations A2-2 and A2-3 in DCC(1,1) and ADCC(1,1), respectively. From Equation A2-6b we get:

$\hat{E}_{t+1}|\Psi_t \sim N(0, \hat{\Omega}_{t+1})$

$\Rightarrow B_{r_t}^{\mathcal{P}\ell} \hat{E}_{t+1}^{\ell}|\Psi_t \sim N(0, \hat{\Omega}_{t+1})$

$\Rightarrow \hat{E}_{t+1}^{\ell}|\Psi_t \sim N\left(0, B_{r_t}^{\mathcal{P}\ell^{-1}} \hat{\Omega}_{t+1}\right)$

We thus establish the following liquidity adjusted DCC/ADCC(1,1) under extreme liquidity:

$$\hat{E}_{t+1}^{\ell}|\Psi_t \sim N(0, \hat{\Omega}_{t+1}^{\ell}) \tag{A2-8a}$$

$$\text{where: } \hat{\Omega}_{t+1}^{\ell} = B_{r_t}^{\mathcal{P}\ell^{-1}} \hat{\Omega}_{t+1} \Rightarrow \hat{\Omega}_{t+1} = B_{r_t}^{\mathcal{P}\ell} \hat{\Omega}_{t+1}^{\ell}$$

Taking advantage of that $B_{r_t}^{\mathcal{P}\ell}$ is a diagonal matrix, from the above equation we get:

$$\hat{\Omega}_{t+1} = B_{r_t}^{\mathcal{P}\ell\frac{1}{2}} B_{r_t}^{\mathcal{P}\ell\frac{1}{2}} \hat{\Omega}_{t+1}^{\ell}$$

$$\text{where: } B_{r_t}^{\mathcal{P}\ell\frac{1}{2}} \text{ is a diagonal matrix}$$

Furthermore, as $\hat{\Omega}_{t+1}^{\ell}$ and $\hat{\Omega}_{t+1}$ are conditional covariance matrices and therefore symmetric and that $B_{r_t}^{\mathcal{P}\ell\frac{1}{2}}$ is a diagonal matrix, and therefore $B_{r_t}^{\mathcal{P}\ell\frac{1}{2}} \hat{\Omega}_{t+1}^{\ell}$ is symmetric, we derive the follows:

$$B_{r_t}^{\mathcal{P}\ell\frac{1}{2}} \hat{\Omega}_{t+1}^{\ell} = {B_{r_t}^{\mathcal{P}\ell\frac{1}{2}}}^H {\hat{\Omega}_{t+1}^{\ell}}^H = \left(\hat{\Omega}_{t+1}^{\ell} B_{r_t}^{\mathcal{P}\ell\frac{1}{2}}\right)^H = \hat{\Omega}_{t+1}^{\ell} B_{r_t}^{\mathcal{P}\ell\frac{1}{2}}$$

$$\Rightarrow \hat{\Omega}_{t+1} = B_{r_t}^{\mathcal{P}\ell\frac{1}{2}} B_{r_t}^{\mathcal{P}\ell\frac{1}{2}} \hat{\Omega}_{t+1}^{\ell} = B_{r_t}^{\mathcal{P}\ell\frac{1}{2}} \left(B_{r_t}^{\mathcal{P}\ell\frac{1}{2}} \hat{\Omega}_{t+1}^{\ell}\right) = B_{r_t}^{\mathcal{P}\ell\frac{1}{2}} \hat{\Omega}_{t+1}^{\ell} B_{r_t}^{\mathcal{P}\ell\frac{1}{2}} \Rightarrow \hat{\Omega}_{t+1}^{\ell} = B_{r_t}^{\mathcal{P}\ell-\frac{1}{2}} \hat{\Omega}_{t+1} B_{r_t}^{\mathcal{P}\ell-\frac{1}{2}} \tag{A2-8b}$$

Also, Equation 7 gives the connection between the daily regular and liquidity-adjusted covariance matrices, and with substitution of scripts we have:

$$\Sigma_{r_t}^{TT} = B_{\sigma_t}^{\mathcal{P}\ell} \Sigma_{r_t^{\ell}}^{TT} B_{\sigma_t}^{\mathcal{P}\ell H} \Rightarrow \Sigma_{r_t^{\ell}}^{TT} = B_{\sigma_t}^{\mathcal{P}\ell^{-1}} \Sigma_{r_t}^{TT} B_{\sigma_t}^{\mathcal{P}\ell H^{-1}} \tag{A2-9}$$

Similar as for the expected regular error vector, for each rolling window, we fit both DCC(1,1) and ADCC(1,1) on the expected liquidity-adjusted error vector $\hat{E}_{t+1}^{\ell}$ to produce the conditional covariance matrix $\hat{\Omega}_{t+1}^{\ell}$, and further estimate the posterior daily covariance matrix for day $t+1$, $\hat{\Sigma}_{r_{t+1}^{\ell}}^{TT}$, which is analytically expressed as:

$$\hat{\Sigma}_{r_{t+1}^{\ell}}^{TT} = \Sigma_{r_t^{\ell}}^{TT} + \left[\left(\tau \Sigma_{r_t^{\ell}}^{TT}\right)^{-1} + \hat{\Omega}_{t+1}^{\ell}{}^{-1}\right]^{-1} \tag{A2-10}$$

Consolidating the above equations, we get the liquidity-adjusted VECM-DCC/ADCC model and posterior daily covariance matrix at extreme liquidity as:

$$Q_t^\ell = \sum_{i=1}^{p} \Phi_i Q_{t-i}^\ell + E_t^\ell \tag{A2-6a}$$

$$\hat{E}_{t+1}^\ell = \hat{Q}_{t+1}^\ell - Q_{t+1}^\ell \tag{A2-7}$$

$$\hat{E}_{t+1}^\ell | \Psi_t \sim N(0, \hat{\Omega}_{t+1}^\ell) \tag{A2-8a}$$

$$\hat{\Sigma}_{r_{t+1}^\ell}^{TT} = \Sigma_{r_t^\ell}^{TT} + \left[\left(\tau \Sigma_{r_t^\ell}^{TT}\right)^{-1} + \hat{\Omega}_{t+1}^{\ell\,-1}\right]^{-1} \tag{A2-10}$$

*where*:

$$Q_t = B_{r_t}^{\mathcal{P}\ell} Q_t^\ell \Rightarrow Q_t^\ell = B_{r_t}^{\mathcal{P}\ell\,-1} Q_t \tag{A2-5}$$

$$E_t^\ell = B_{r_t}^{\mathcal{P}\ell\,-1} E_t \tag{A2-6b}$$

$$\Sigma_{r_t^\ell}^{TT} = B_{\sigma_t}^{\mathcal{P}\ell\,-1} \Sigma_{r_t}^{TT} B_{\sigma_t}^{\mathcal{P}\ell\,H\,-1} \tag{A2-9}$$

$$\hat{\Omega}_{t+1}^\ell = B_{r_t}^{\mathcal{P}\ell\,-\frac{1}{2}} \hat{\Omega}_{t+1} B_{r_t}^{\mathcal{P}\ell\,-\frac{1}{2}} \tag{A2-8b}$$

We then transfer the above equation block back to Subsection 5.1 as Equations 12-19.

### A2.3 Linkage and Comparisons of Posterior Daily Covariance Matrices

In addition to Equations A2-5, A2-6b, A2-9 and A2-8b, we conduct the following matrix manipulations in order to connect $\hat{\Sigma}_{r_{t+1}^\ell}^{TT}$ with $\hat{\Sigma}_{r_{t+1}}^{TT}$:

$$\hat{\Sigma}_{r_{t+1}^\ell}^{TT} = \Sigma_{r_t^\ell}^{TT} + \left[\left(\tau \Sigma_{r_t^\ell}^{TT}\right)^{-1} + \hat{\Omega}_{t+1}^{\ell\,-1}\right]^{-1} \tag{A2-10}$$

$$\Rightarrow \hat{\Sigma}_{r_{t+1}^\ell}^{TT} = B_{\sigma_t}^{\mathcal{P}\ell\,-1} \Sigma_{r_t}^{TT} B_{\sigma_t}^{\mathcal{P}\ell\,H\,-1} + \left[\left(\tau B_{\sigma_t}^{\mathcal{P}\ell\,-1} \Sigma_{r_t}^{TT} B_{\sigma_t}^{\mathcal{P}\ell\,H\,-1}\right)^{-1} + \left(B_{r_t}^{\mathcal{P}\ell\,-\frac{1}{2}} \hat{\Omega}_{t+1} B_{r_t}^{\mathcal{P}\ell\,-\frac{1}{2}}\right)^{-1}\right]^{-1}$$

$$\Rightarrow \hat{\Sigma}_{r_{t+1}^\ell}^{TT} = B_{\sigma_t}^{\mathcal{P}\ell\,-1} \Sigma_{r_t}^{TT} B_{\sigma_t}^{\mathcal{P}\ell\,H\,-1} + \left[B_{\sigma_t}^{\mathcal{P}\ell\,H}\left(\tau \Sigma_{r_t}^{TT}\right)^{-1} B_{\sigma_t}^{\mathcal{P}\ell} + B_{r_t}^{\mathcal{P}\ell\,\frac{1}{2}} \hat{\Omega}_{t+1}^{-1} B_{r_t}^{\mathcal{P}\ell\,\frac{1}{2}}\right]^{-1}$$

$$\Rightarrow \hat{\Sigma}_{r_{t+1}^\ell}^{TT} = B_{\sigma_t}^{\mathcal{P}\ell\,-1} \Sigma_{r_t}^{TT} B_{\sigma_t}^{\mathcal{P}\ell\,H\,-1} + \left[B_{\sigma_t}^{\mathcal{P}\ell\,H}\left(\tau \Sigma_{r_t}^{TT}\right)^{-1} B_{\sigma_t}^{\mathcal{P}\ell} + B_{\sigma_t}^{\mathcal{P}\ell\,H} B_{\sigma_t}^{\mathcal{P}\ell\,H\,-1} \left(B_{r_t}^{\mathcal{P}\ell\,\frac{1}{2}} \hat{\Omega}_{t+1}^{-1} B_{r_t}^{\mathcal{P}\ell\,\frac{1}{2}}\right) B_{\sigma_t}^{\mathcal{P}\ell\,-1} B_{\sigma_t}^{\mathcal{P}\ell}\right]^{-1}$$

$$\Rightarrow \hat{\Sigma}_{r_{t+1}^\ell}^{TT} = B_{\sigma_t}^{\mathcal{P}\ell\,-1} \Sigma_{r_t}^{TT} B_{\sigma_t}^{\mathcal{P}\ell\,H\,-1} + \left[B_{\sigma_t}^{\mathcal{P}\ell\,H}\left[\left(\tau \Sigma_{r_t}^{TT}\right)^{-1} + B_{\sigma_t}^{\mathcal{P}\ell\,H\,-1}\left(B_{r_t}^{\mathcal{P}\ell\,\frac{1}{2}} \hat{\Omega}_{t+1}^{-1} B_{r_t}^{\mathcal{P}\ell\,\frac{1}{2}}\right) B_{\sigma_t}^{\mathcal{P}\ell\,-1}\right] B_{\sigma_t}^{\mathcal{P}\ell}\right]^{-1}$$

$$\Rightarrow \hat{\Sigma}_{r_{t+1}^\ell}^{TT} = B_{\sigma_t}^{\mathcal{P}\ell\,-1} \Sigma_{r_t}^{TT} B_{\sigma_t}^{\mathcal{P}\ell\,H\,-1} + B_{\sigma_t}^{\mathcal{P}\ell\,-1}\left[\left(\tau \Sigma_{r_t}^{TT}\right)^{-1} + \left(B_{\sigma_t}^{\mathcal{P}\ell\,H\,-1} B_{r_t}^{\mathcal{P}\ell\,\frac{1}{2}}\right) \hat{\Omega}_{t+1}^{-1} \left(B_{r_t}^{\mathcal{P}\ell\,\frac{1}{2}} B_{\sigma_t}^{\mathcal{P}\ell\,-1}\right)\right]^{-1} B_{\sigma_t}^{\mathcal{P}\ell\,H\,-1}$$

$$\Rightarrow \hat{\Sigma}_{r_{t+1}^\ell}^{TT} = B_{\sigma_t}^{\mathcal{P}\ell\,-1} \Sigma_{r_t}^{TT} B_{\sigma_t}^{\mathcal{P}\ell\,H\,-1} + B_{\sigma_t}^{\mathcal{P}\ell\,-1}\left[\left(\tau \Sigma_{r_t}^{TT}\right)^{-1} + \left(B_{r_t}^{\mathcal{P}\ell\,\frac{1}{2}} B_{\sigma_t}^{\mathcal{P}\ell\,-1}\right)^{H} \hat{\Omega}_{t+1}^{-1} \left(B_{r_t}^{\mathcal{P}\ell\,\frac{1}{2}} B_{\sigma_t}^{\mathcal{P}\ell\,-1}\right)\right]^{-1} B_{\sigma_t}^{\mathcal{P}\ell\,H\,-1}$$

$$\Rightarrow \hat{\Sigma}_{r_{t+1}^{\ell}}^{TT} = B_{\sigma_t}^{\mathcal{P}\ell-1}\Sigma_{r_t}^{TT}B_{\sigma_t}^{\mathcal{P}\ell\,H-1} + B_{\sigma_t}^{\mathcal{P}\ell-1}\left[\left(\tau\Sigma_{r_t}^{TT}\right)^{-1} + \left(\left(B_{r_t}^{\mathcal{P}\ell\frac{1}{2}}B_{\sigma_t}^{\mathcal{P}\ell-1}\right)^{-1}\hat{\Omega}_{t+1}\left(B_{r_t}^{\mathcal{P}\ell\frac{1}{2}}B_{\sigma_t}^{\mathcal{P}\ell-1}\right)^{H-1}\right)^{-1}\right]^{-1}B_{\sigma_t}^{\mathcal{P}\ell\,H-1}$$

$$\Rightarrow \hat{\Sigma}_{r_{t+1}^{\ell}}^{TT} = B_{\sigma_t}^{\mathcal{P}\ell-1}\Sigma_{r_t}^{TT}B_{\sigma_t}^{\mathcal{P}\ell\,H-1} + B_{\sigma_t}^{\mathcal{P}\ell-1}\left[\left(\tau\Sigma_{r_t}^{TT}\right)^{-1} + \left(\left(B_{r_t}^{\mathcal{P}\ell\frac{1}{2}}B_{\sigma_t}^{\mathcal{P}\ell-1}\right)^{-1}\hat{\Omega}_{t+1}\left(B_{r_t}^{\mathcal{P}\ell\frac{1}{2}}B_{\sigma_t}^{\mathcal{P}\ell-1}\right)^{-1\,H}\right)^{-1}\right]^{-1}B_{\sigma_t}^{\mathcal{P}\ell\,H-1}$$

$$\Rightarrow \hat{\Sigma}_{r_{t+1}^{\ell}}^{TT} = B_{\sigma_t}^{\mathcal{P}\ell-1}\Sigma_{r_t}^{TT}B_{\sigma_t}^{\mathcal{P}\ell\,H-1} + B_{\sigma_t}^{\mathcal{P}\ell-1}\left[\left(\tau\Sigma_{r_t}^{TT}\right)^{-1} + \left(\left(B_{\sigma_t}^{\mathcal{P}\ell}B_{r_t}^{\mathcal{P}\ell-\frac{1}{2}}\right)\hat{\Omega}_{t+1}\left(B_{\sigma_t}^{\mathcal{P}\ell}B_{r_t}^{\mathcal{P}\ell-\frac{1}{2}}\right)^{H}\right)^{-1}\right]^{-1}B_{\sigma_t}^{\mathcal{P}\ell\,H-1}$$

$$\Rightarrow \hat{\Sigma}_{r_{t+1}^{\ell}}^{TT} = B_{\sigma_t}^{\mathcal{P}\ell-1}\Sigma_{r_t}^{TT}B_{\sigma_t}^{\mathcal{P}\ell\,H-1} + B_{\sigma_t}^{\mathcal{P}\ell-1}\left[\left(\tau\Sigma_{r_t}^{TT}\right)^{-1} + \left(B_t^{\mathcal{P}\ell}\hat{\Omega}_{t+1}B_t^{\mathcal{P}\ell\,H}\right)^{-1}\right]^{-1}B_{\sigma_t}^{\mathcal{P}\ell\,H-1}$$

$$\Rightarrow \hat{\Sigma}_{r_{t+1}^{\ell}}^{TT} = B_{\sigma_t}^{\mathcal{P}\ell-1}\left[\Sigma_{r_t}^{TT} + \left[\left(\tau\Sigma_{r_t}^{TT}\right)^{-1} + \left(B_t^{\mathcal{P}\ell}\hat{\Omega}_{t+1}B_t^{\mathcal{P}\ell\,H}\right)^{-1}\right]^{-1}\right]B_{\sigma_t}^{\mathcal{P}\ell\,H-1} \qquad (A2\text{-}11)$$

$$\text{where: } B_t^{\mathcal{P}\ell} = B_{\sigma_t}^{\mathcal{P}\ell}B_{r_t}^{\mathcal{P}\ell-\frac{1}{2}} \qquad (A2\text{-}12)$$

We then transfer Equations A2-11 and A2-12 back to Subsection 5.3 as Equations 20 and 21.

# References

Al Janabi, M.A., 2011. Dynamic equity asset allocation with liquidity-adjusted market risk criterion: Appraisal of efficient and coherent portfolios. *J. of Asset Management*, *12*, pp.378-394.

Al Janabi, M.A., 2013. Optimal and coherent economic-capital structures: evidence from long and short-sales trading positions under illiquid market perspectives. *Annals of Operations Research*, *205*(1), pp.109-139.

Al Janabi, M.A., 2021. Multivariate portfolio optimization under illiquid market prospects: a review of theoretical algorithms and practical techniques for liquidity risk management. *Journal of Modelling in Management*, *16*(1), pp.288-309.

Al Janabi, M.A., Ferrer, R. and Shahzad, S.J.H., 2019. Liquidity-adjusted value-at-risk optimization of a multi-asset portfolio using a vine copula approach. *Physica A: Statistical Mechanics and its Applications*, *536*, p.122579.

Al Janabi, M.A., Hernandez, J.A., Berger, T. and Nguyen, D.K., 2017. Multivariate dependence and portfolio optimization algorithms under illiquid market scenarios. *European Journal of Operational Research*, *259*(3), pp.1121-1131.

Brandt, M.W., Goyal, A., Santa-Clara, P. and Stroud, J.R., 2005. A simulation approach to dynamic portfolio choice with an application to learning about return predictability. *The Review of Financial Studies*, *18*(3), pp.831-873.

Cappiello, L., Engle, R.F. and Sheppard, K., 2006. Asymmetric dynamics in the correlations of global equity and bond returns. *Journal of Financial Econometrics 4*(4), pp.537-572.

Çetin, U. and Rogers, L.C.G., 2007. Modeling liquidity effects in discrete time. *Mathematical Finance*, *17*(1), pp.15-29.

Chong, J. and Miffre, J., 2010. Conditional correlation and volatility in commodity futures and traditional asset markets. *Journal of Alternative Investments*, *12*(3), p.61.

Cong, F. and Oosterlee, C.W., 2016. Multi-period mean–variance portfolio optimization based on Monte-Carlo simulation. *Journal of Economic Dynamics and Control*, *64*, pp.23-38.

Cong, F. and Oosterlee, C.W., 2017. Accurate and robust numerical methods for the dynamic portfolio management problem. *Computational Economics*, *49*, pp.433-458.

Cong, L. W., Li, X., Tang, K., and Yang, Y., 2023. Crypto wash trading. *Management Science* 69, pp.6427-6454.

Deng, Q., 2018. A generalized VECM/VAR-DCC/ADCC framework and its application in the Black-Litterman model: Illustrated with a China portfolio. *China Finance Review International*, *8*(4), pp.453-467.

Deng, Q., 2024. A conditional singular value decomposition. Available at arXiv: https://doi.org/10.48550/arXiv.2403.09696, or at SSRN: https://dx.doi.org/10.2139/ssrn.4736134.

Deng, Q. and Zhou, Z.G., 2024. Liquidity Jump, Liquidity Diffusion, and Crypto Wash Trading. Available at arXiv: https://doi.org/10.48550/arXiv.2411.05803, or at SSRN: https://dx.doi.org/10.2139/ssrn.5001541.

Deng, Q. and Zhou, Z.G., 2025. Liquidity-adjusted Return and Volatility, and Autoregressive Models. Available at arXiv: https://doi.org/10.48550/arXiv.2503.08693, or at SSRN: https://dx.doi.org/10.2139/ssrn.5161751.

Engle, R.F. and Granger, C.W., 1987. Co-integration and error correction: representation, estimation, and testing. *Econometrica: Journal of the Econometric Society*, *55*(2), pp.251-276.

Engle, R.F. and Sheppard, K., 2001. Theoretical and empirical properties of dynamic conditional correlation multivariate GARCH. NBER Working Paper 8554.

Gaigi, M.H., Ly Vath, V., Mnif, M. and Toumi, S., 2016. Numerical approximation for a portfolio optimization problem under liquidity risk and costs. *Applied Math & Op*, *74*, pp.163-195.

Gârleanu, N. and Pedersen, L.H., 2013. Dynamic trading with predictable returns and transaction costs. *The Journal of Finance*, *68*(6), pp.2309-2340.

Hung, J.C., Su, J.B., Chang, M.C. and Wang, Y.H., 2020. The impact of liquidity on portfolio value-at-risk forecasts. *Applied economics*, *52*(3), pp.242-259.

Kolm, P. N., Tütüncü, R., and Fabozzi, F. J., 2014. 60 years of portfolio optimization: Practical challenges and current trends. *European Journal of Operational Research, 234(2)*, pp.356-371.

Lim, A.E. and Wimonkittiwat, P., 2014. Dynamic portfolio selection with market impact costs. *Operations Research Letters*, *42*(5), pp.299-306.

Ling, S. and McAleer, M., 2003. Asymptotic theory for a vector ARMA-GARCH model. *Econometric Theory*, *19*(2), pp.280-310.

Lo, A.W., Petrov, C. and Wierzbicki, M., 2006. It's 11 Pm—Do you know where your liquidity is? The mean–variance–liquidity frontier. In *The World of Risk Management*, pp.47-92.

Ly Vath, V., Mnif, M. and Pham, H., 2007. A model of optimal portfolio selection under liquidity risk and price impact. *Finance and Stochastics*, *11*, pp.51-90.

Ma, J., Song, Q., Xu, J. and Zhang, J., 2013. Optimal portfolio selection under concave price impact. *Applied Mathematics & Optimization*, *67*(3), pp.353-390.

Mei, X., DeMiguel, V. and Nogales, F.J., 2016. Multiperiod portfolio optimization with multiple risky assets and general transaction costs. *Journal of Banking & Finance*, *69*, pp.108-120.

Phillips, P.C. and Ouliaris, S., 1990. Asymptotic properties of residual based tests for cointegration. *Econometrica: journal of the Econometric Society*, *58*(1), pp.165-193.

Vieira, E.B.F. and Filomena, T.P., 2020. Liquidity constraints for portfolio selection based on financial volume. *Computational Economics*, *56*(4), pp.1055-1077.

Vieira, E.B.F., Filomena, T.P., Sant'anna, L.R. and Lejeune, M.A., 2023. Liquidity-constrained index tracking optimization models. *Annals of Operations Research*, *330*(1), pp.73-118.

Weiß, G.N. and Supper, H., 2013. Forecasting liquidity-adjusted intraday value-at-risk with vine copulas. *Journal of Banking & Finance*, *37*(9), pp.3334-3350.

Zhang, R., Langrené, N., Tian, Y., Zhu, Z., Klebaner, F. and Hamza, K., 2019. Dynamic portfolio optimization with liquidity cost and market impact: a simulation-and-regression approach. *Quantitative Finance*, *19*(3), pp.519-532.

# Table 1 – Descriptive Statistics of Portfolio Liquidity Jump $|B_{r_t}^{\mathcal{P}\ell}|$, Portfolio Liquidity Diffusion $|B_{\sigma_t}^{\mathcal{P}\ell}|$, and Portfolio Liquidity $|B_t^{\mathcal{P}\ell}|$

Panel A reports descriptive statistics of portfolio liquidity jump $|B_{r_t}^{\mathcal{P}\ell}|$, portfolio liquidity diffusion $|B_{\sigma_t}^{\mathcal{P}\ell}|$, and the composite portfolio liquidity $|B_t^{\mathcal{P}\ell}|$ of a portfolio comprising all eight cryptocurrencies over the entire sampling period. Panel B reports descriptive statistics of portfolio liquidity jump $|B_{r_t}^{\mathcal{P}\ell}|$ and portfolio liquidity diffusion $|B_{\sigma_t}^{\mathcal{P}\ell}|$ of a portfolio comprising all 15 US stocks over the entire sampling period. The maximum values of $|B_{r_t}^{\mathcal{P}\ell}|$, $|B_{\sigma_t}^{\mathcal{P}\ell}|$ and $B_t^{\mathcal{P}\ell}$ are capped at 10. The composite portfolio liquidity $|B_t^{\mathcal{P}\ell}|$ is defined as: $B_t^{\mathcal{P}\ell} = B_{\sigma_t}^{\mathcal{P}\ell} B_{r_t}^{\mathcal{P}\ell - \frac{1}{2}}$ (21).

| Panel A | Cryptocurrency Portfolio | | |
|---|---|---|---|
| measures | liquidity jump $|B_{r_t}^{\mathcal{P}\ell}|$ | liquidity diffusion $|B_{\sigma_t}^{\mathcal{P}\ell}|$ | liquidity composite $|B_t^{\mathcal{P}\ell}|$ |
| count | 2242 | 2242 | 2242 |
| mean | 3.77 | 1.38 | 2.59 |
| std | 4.45 | 2.56 | 3.51 |
| min | 0.00 | 0.01 | 0.01 |
| median | 0.70 | 0.33 | 0.68 |
| max | 10.00 | 10.00 | 10.00 |
| number of days (= 10) | 682 | 204 | 312 |
| as % of total number of days | 30.42% | 23.42% | 13.92% |
| number of days (>= 1) | 1051 | 811 | 987 |
| as % of total number of days | 46.88% | 36.17% | 44.02% |
| number of days (<= 0.10) | 748 | 3 | 340 |
| as % of total number of days | 33.36% | 0.13% | 15.17% |

| Panel B | Stock Portfolio | | |
|---|---|---|---|
| measures | liquidity jump $|B_{r_t}^{\mathcal{P}\ell}|$ | liquidity diffusion $|B_{\sigma_t}^{\mathcal{P}\ell}|$ | liquidity composite $|B_t^{\mathcal{P}\ell}|$ |
| count | 2671 | 2671 | 2671 |
| mean | 0.11 | 0.00 | 0.72 |
| std | 0.86 | 0.00 | 2.39 |
| min | 0.00 | 0.00 | 0.00 |
| median | 0.00 | 0.00 | 0.01 |
| max | 10 | 0.00 | 10.00 |
| number of days (= 10) | 16 | 0 | 157 |
| as % of total number of days | 0.60% | 0.00% | 5.88% |
| number of days (>= 1) | 48 | 0 | 240 |
| as % of total number of days | 1.80% | 0.00% | 8.99% |
| number of days (<= 0.10) | 2531 | 2671 | 2086 |
| as % of total number of days | 94.76% | 100.00% | 78.10% |

Notes:

1. The "number of days (=10)," "number of days (>=1)," and "number of days (<=0.10)" rows give the numbers of trading days under that particular condition for the liquidity measures. The "as % of total number of trading days" row under each of the above gives the number of trading days under that particular condition as a percentage of the total number of trading days (2,242 for cryptocurrencies or 2,671 for stocks).

## Table 2 – One-sided *t*-tests for Conditional Covariance and Posterior Variance

Panel A reports the one-sided *t*-test results for the determinant of regular conditional covariance $|\hat{\Omega}_{t+1}|$ and the determinant of liquidity-adjusted conditional covariance $|\hat{\Omega}^{\ell}_{t+1}|$ for both cryptocurrency portfolio and stock portfolio. Panel B reports the one-sided *t*-test results for the determinant of regular posterior covariance $|\hat{\Sigma}^{TT}_{r_{t+1}}|$ and the determinant of liquidity-adjusted posterior covariance $|\hat{\Sigma}^{TT}_{r^{\ell}_{t+1}}|$ for both cryptocurrency portfolio and stock portfolio.

| Panel A: conditional covariance | | one-sided *t*-test | | alternative hypothesis | *t*-value | dof | p_value greater | p_value less | sig | interpretation | direction |
|---|---|---|---|---|---|---|---|---|---|---|---|
| cryptocurrency portfolio | dcc *t*-test | $|\hat{\Omega}_{t+1}| - |\hat{\Omega}^{\ell}_{t+1}|$ | < 0 | liquidity adjustment increases conditional covariance | -1.87 | 3752 | 0.97 | **0.03** | ** | liquidity adjustment increases conditional covariance | ↑ |
| | adcc *t*-test | | | | -1.83 | 3752 | 0.97 | **0.03** | ** | | ↑ |
| | dcc_best *t*-test | | | | -1.83 | 3752 | 0.97 | **0.03** | ** | | ↑ |
| stock portfolio | dcc *t*-test | $|\hat{\Omega}_{t+1}| - |\hat{\Omega}^{\ell}_{t+1}|$ | < 0 | liquidity adjustment increases conditional covariance | -1.33 | 4856 | 0.91 | **0.09** | * | liquidity adjustment increases conditional covariance | ↑ |
| | adcc *t*-test | | | | -1.41 | 4856 | 0.92 | **0.08** | * | | ↑ |
| | dcc_best *t*-test | | | | -1.33 | 4856 | 0.91 | **0.09** | * | | ↑ |

| Panel B: posterior covariance | | one-sided *t*-test | | alternative hypothesis | *t*-value | dof | p_value greater | p_value less | sig | interpretation | direction |
|---|---|---|---|---|---|---|---|---|---|---|---|
| cryptocurrency portfolio | dcc *t*-test | $|\hat{\Sigma}^{TT}_{r_{t+1}}| - |\hat{\Sigma}^{TT}_{r^{\ell}_{t+1}}|$ | < 0 | liquidity adjustment increases posterior covariance | -1.38 | 3752 | 0.92 | **0.08** | * | liquidity adjustment increases posterior covariance | ↑ |
| | adcc *t*-test | | | | -1.35 | 3752 | 0.91 | **0.09** | * | | ↑ |
| | dcc_best *t*-test | | | | -1.35 | 3752 | 0.91 | **0.09** | * | | ↑ |
| stock portfolio | dcc *t*-test | $|\hat{\Sigma}^{TT}_{r_{t+1}}| - |\hat{\Sigma}^{TT}_{r^{\ell}_{t+1}}|$ | < 0 | liquidity adjustment increases posterior covariance | -1.00 | 4856 | 0.84 | **0.16** | | no significant evidence that liquidity adjustment increases posterior covariance | ↔ |
| | adcc *t*-test | | | | -1.00 | 4856 | 0.84 | **0.16** | | | ↔ |
| | dcc_best *t*-test | | | | -1.00 | 4856 | 0.84 | **0.16** | | | ↔ |

\*\*\* - significant at 1% level, \*\* - significant at 5% level, \* significant at 10% level.

Notes:

1. The terms "dcc," "adcc" and "dcc_best" specify the DCC specifications in deriving the conditional and posterior covariance matrices. The term "dcc" refers to the DCC(1,1) specification in Equation A2-3a; the term "adcc" refers to the ADCC(1,1) specification in Equation A2-3b; and the term "dcc_best" refers to either DCC(1,1) or ADCC(1,1) with a higher log-likelihood for the specific daily data on day *t*.

2. The "direction" column indicates the effect of liquidity adjustment: ↑ indicates that liquidity adjustment increases the value of the tested variable, ↓ means liquidity adjustment reduces the value of the test variable, and ↔ indicates liquidity adjustment has no statistically significant impact on the test variable.

## Table 3 – Two-sided $t$-tests for Coefficients of DCC and ADCC

Panel A reports the two-sided $t$-test results for the coefficients ($a$, $b$, $g$) of DCC(1,1) specification (Equation A2-3a) and ADCC(1,1) specification (Equation A2-3b) for the regular and liquidity-adjusted conditional variance estimations for the cryptocurrency portfolio. Panel B reports the same for the stock portfolio.

$$\hat{E}_{t+1}|\Psi_t \sim N(0, \hat{\Omega}_{t+1} = \hat{H}_{t+1}\hat{P}_{t+1}\hat{H}_{t+1}) \quad (A2\text{-}3a)$$
$$\hat{H}_{t+1}^2 = H_0^2 + KE_t E_t^H + \Lambda H_t^2$$
$$\hat{P}_{t+1} = \hat{O}_{t+1}^* \hat{O}_{t+1} \hat{O}_{t+1}^*$$
$$\hat{O}_{t+1} = (1 - a - b)\bar{O} + a\Xi_t \Xi_t^H + bO_t$$
$$\Xi_t = H_t^{-1} E_t$$
$$a + b < 1$$

$$\hat{E}_{t+1}|\Psi_t \sim N(0, \hat{\Omega}_{t+1} = \hat{H}_{t+1}\hat{P}_{t+1}\hat{H}_{t+1}) \quad (A2\text{-}3b)$$
$$\hat{H}_{t+1}^2 = H_0^2 + KE_t E_t^H + \Lambda H_t^2$$
$$\hat{P}_{t+1} = \hat{O}_{t+1}^* \hat{O}_{t+1} \hat{O}_{t+1}^*$$
$$\hat{O}_{t+1} = (1 - a - b)\bar{O} - g\bar{N} + a\Xi_t \Xi_t^H + bO_t + gN_t N_t^H$$
$$\Xi_t = H_t^{-1} E_t$$
$$N_t = I[\xi_{i,t} < 0] \circ \Xi_t$$
$$a + b + g < 1$$

| | Panel A: cryptocurrency portfolio | $t$-value | dof | $p$-value two-sided | $p$-value greater | $p$-value less | sig | interpretation | direction |
|---|---|---|---|---|---|---|---|---|---|
| dcc $t$-test (two-sided) | shock sensitive coefficient $a$ | 4.86 | 3752 | 0.00 | **0.00** | 1.00 | *** | liquidity adjustment reduces $a$ | ↓ |
| | correlation persistence coefficient $b$ | 13.64 | 3752 | 0.00 | **0.00** | 1.00 | *** | liquidity adjustment reduces $b$ | ↓ |
| | $a + b$ | 14.06 | 3752 | 0.00 | **0.00** | 1.00 | *** | liquidity adjustment reduces $a + b$ | ↓ |
| adcc $t$-test (two-sided) | shock sensitive coefficient $a$ | -0.07 | 3752 | **0.94** | 0.53 | 0.47 | | liquidity adjustment has no significant impact on $a$ | ↔ |
| | correlation persistence coefficient $b$ | 17.84 | 3752 | 0.00 | **0.00** | 1.00 | *** | liquidity adjustment reduces $b$ | ↓ |
| | negative shock sensitive coefficient $g$ | 17.32 | 3752 | 0.00 | **0.00** | 1.00 | *** | liquidity adjustment reduces $g$ | ↓ |
| | $a + b + g$ | 21.19 | 3752 | 0.00 | **0.00** | 1.00 | *** | liquidity adjustment reduces $a + b + g$ | ↓ |

| | Panel B: stock portfolio | $t$-value | dof | $p$-value two-sided | $p$-value greater | $p$-value less | sig | interpretation | direction |
|---|---|---|---|---|---|---|---|---|---|
| dcc $t$-test (two-sided) | shock sensitive coefficient $a$ | 12.34 | 4856 | 0.00 | **0.00** | 1.00 | *** | liquidity adjustment reduces $a$ | ↓ |
| | correlation persistence coefficient $b$ | -12.80 | 4856 | 0.00 | 1.00 | **0.00** | *** | liquidity adjustment increases $b$ | ↑ |
| | $a + b$ | -12.76 | 4856 | 0.00 | 1.00 | **0.00** | *** | liquidity adjustment increases $a + b$ | ↑ |
| adcc $t$-test (two-sided) | shock sensitive coefficient $a$ | 11.80 | 4856 | 0.00 | **0.00** | 1.00 | *** | liquidity adjustment reduces $a$ | ↓ |
| | correlation persistence coefficient $b$ | -17.46 | 4856 | 0.00 | 1.00 | **0.00** | *** | liquidity adjustment increases $b$ | ↑ |
| | negative shock sensitive coefficient $g$ | 13.48 | 4856 | 0.00 | **0.00** | 1.00 | *** | liquidity adjustment reduces $g$ | ↓ |
| | $a + b + g$ | -17.35 | 4856 | 0.00 | 1.00 | **0.00** | *** | liquidity adjustment increases $a + b + g$ | ↑ |

\*\*\* - significant at 1% level, \*\* - significant at 5% level, \* significant at 10% level.

Notes:

1. The terms "dcc" and "adcc" specify the DCC specifications in deriving the conditional covariance matrix. The term "dcc" refers to the DCC(1,1) specification in Equation A2-3a; the term "adcc" refers to the ADCC(1,1) specification in Equation A2-3b. Both equations are listed below.

2. The "direction" column indicates the effect of liquidity adjustment: ↑ indicates that liquidity adjustment increases the value of the tested variable, ↓ means liquidity adjustment reduces the value of the test variable, and ↔ indicates liquidity adjustment has no statistically significant impact on the test variable.

## Table 4 – Performance Comparisons of TMV vs. LAMV

Panel A compares the performance of six portfolios, including three TMV portfolios (1, 3, 5) and three LAMV portfolios (2, 4, 6) for the cryptocurrency portfolios. Panel B compares the performance of six portfolios, including three TMV portfolios (1, 3, 5) and three LAMV portfolios (2, 4, 6) for the US stock portfolios.

| Panel A | Cryptocurrency Portfolios | |
|---|---|---|
| Portfolio Number | 1 | 2 |
| Portfolio Description | standard TMV | standard LAMV |
| Return in MV | regular rolling window $\bar{\mu}_{r_t}$ | liquidity-adjusted rolling window $\bar{\mu}_{r_t^\ell}$ |
| Covariance in MV | regular rolling window $\bar{\Sigma}_{r_t}$ | liquidity-adjusted rolling window $\bar{\Sigma}_{r_t^\ell}$ |
| Annualized Sharpe Ratio ($r_f$ = 0%) | **0.83** | **0.67** |
| Portfolio Number | 3 | 4 |
| Portfolio Description | intraday TMV | intraday LAMV |
| Return in MV | regular rolling window $\bar{\mu}_{r_t}$ | liquidity-adjusted rolling window $\bar{\mu}_{r_t^\ell}$ |
| Covariance in MV | regular intraday $\Sigma_{r_t}^{TT}$ | liquidity-adjusted intraday $\Sigma_{r_t^\ell}^{TT}$ |
| Annualized Sharpe Ratio ($r_f$ = 0%) | **0.52** | **0.94** |
| Portfolio Number | 5 | 6 |
| Portfolio Description | enhanced TMV | enhanced LAMV |
| Return in MV | regular rolling window $\bar{\mu}_{r_t}$ | liquidity-adjusted rolling window $\bar{\mu}_{r_t^\ell}$ |
| Covariance in MV | regular posterior $\hat{\Sigma}_{r_{t+1}}^{TT}$ | liquidity-adjusted posterior $\hat{\Sigma}_{r_{t+1}^\ell}^{TT}$ |
| Annualized Sharpe Ratio ($r_f$ = 0%) | **0.76** | **1.04** |

| Panel B | Stock Portfolios | |
|---|---|---|
| Portfolio Number | 1 | 2 |
| Portfolio Description | standard TMV | standard LAMV |
| Return in MV | regular rolling window $\bar{\mu}_{r_t}$ | liquidity-adjusted rolling window $\bar{\mu}_{r_t^\ell}$ |
| Covariance in MV | regular rolling window $\bar{\Sigma}_{r_t}$ | liquidity-adjusted rolling window $\bar{\Sigma}_{r_t^\ell}$ |
| Annualized Sharpe Ratio ($r_f$ = 0%) | **1.20** | **1.18** |
| Portfolio Number | 3 | 4 |
| Portfolio Description | intraday TMV | intraday LAMV |
| Return in MV | regular rolling window $\bar{\mu}_{r_t}$ | liquidity-adjusted rolling window $\bar{\mu}_{r_t^\ell}$ |
| Covariance in MV | regular intraday $\Sigma_{r_t}^{TT}$ | liquidity-adjusted intraday $\Sigma_{r_t^\ell}^{TT}$ |
| Annualized Sharpe Ratio ($r_f$ = 0%) | **1.27** | **1.30** |
| Portfolio Number | 5 | 6 |
| Portfolio Description | enhanced TMV | enhanced LAMV |
| Return in MV | regular rolling window $\bar{\mu}_{r_t}$ | liquidity-adjusted rolling window $\bar{\mu}_{r_t^\ell}$ |
| Covariance in MV | regular posterior $\hat{\Sigma}_{r_{t+1}}^{TT}$ | liquidity-adjusted posterior $\hat{\Sigma}_{r_{t+1}^\ell}^{TT}$ |
| Annualized Sharpe Ratio ($r_f$ = 0%) | **1.21** | **1.31** |

Notes:

1. The "Return in MV" row specifies the return used in the linear term of the MV construct. The "Covariance in MV" specifies the covariance matrix used in the quadratic term of the MV construct.

2. The risk-free rate ($r_f$) is assumed to be zero in the calculation of the Annualized Sharpe Ratio.

# Figure 1 – Distributions of Portfolio Liquidity Jump $\left|B_{r_t}^{\mathcal{P}\ell}\right|$, Portfolio Liquidity Diffusion $\left|B_{\sigma_t}^{\mathcal{P}\ell}\right|$, and Portfolio Liquidity $\left|B_t^{\mathcal{P}\ell}\right|$

Column A provides the histograms of portfolio liquidity jump $\left|B_{r_t}^{\mathcal{P}\ell}\right|$ (row 1), portfolio liquidity diffusion $\left|B_{\sigma_t}^{\mathcal{P}\ell}\right|$ (row 2), and the composite portfolio liquidity $\left|B_t^{\mathcal{P}\ell}\right|$ (row 3) for the portfolio of all eight cryptocurrencies over the entire sampling period. Column B provides the histograms of the sample portfolio liquidity metrics for the portfolio of all 15 US stocks over the entire sampling period. The maximum values of $\left|B_{r_t}^{\mathcal{P}\ell}\right|$, $\left|B_{r_t}^{\mathcal{P}\ell}\right|$ and $\left|B_t^{\mathcal{P}\ell}\right|$ are capped at 10. The composite portfolio liquidity $\left|B_t^{\mathcal{P}\ell}\right|$ is defined as:

$$B_t^{\mathcal{P}\ell} = B_{\sigma_t}^{\mathcal{P}\ell} B_{r_t}^{\mathcal{P}\ell - \frac{1}{2}} \qquad (21)$$

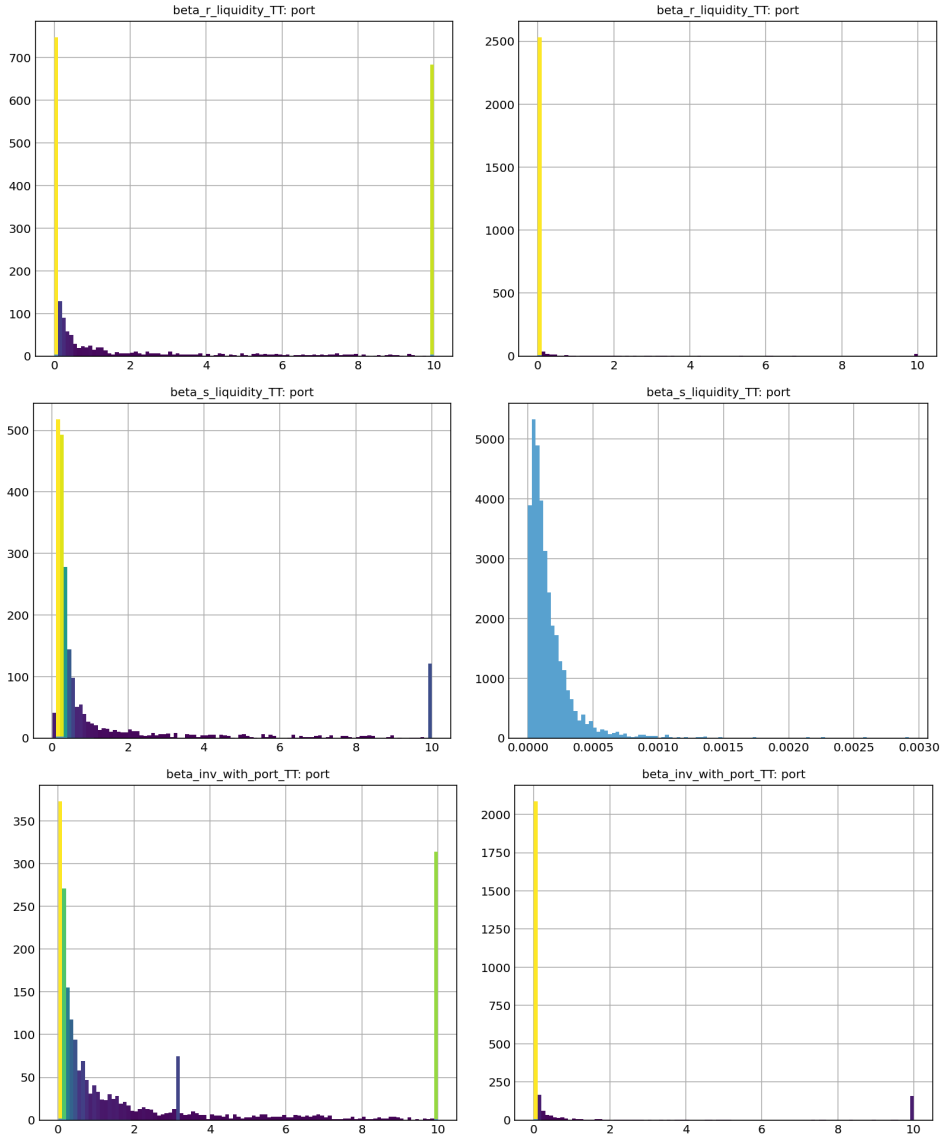